\documentclass[letterpaper, twocolumn, final]{IEEEtran} 
\usepackage{fancyhdr}
\usepackage{epsfig}
\usepackage{threeparttable}
\usepackage{epsf,epsfig}
\usepackage{amsthm}
\usepackage{amsmath}
\usepackage{amssymb}
\usepackage{amsfonts}
\usepackage[noadjust]{cite}
\usepackage{dsfont}
\usepackage{subfigure}
\usepackage{color}
\usepackage{url}
\usepackage{mathrsfs}
\usepackage{multirow}
\usepackage{booktabs}
\usepackage{dashbox}
\usepackage{framed}
\usepackage{multirow}
\usepackage{bigstrut}
\usepackage{array}
\usepackage{setspace}

\newlength{\eqboxstorage}

\newtheorem{algorithm}{Algorithm}

\newtheorem{lemma}{Lemma}

\newtheorem{proposition}{Proposition}

\def\phi{\varphi}

\def\SNR{\mathsf{SNR}}

\def\l{\left}
\def\r{\right}
\def\({\left(}
\def\){\right)}

\def\bff{{\mathbf{f}}}

\def\bh{{\mathbf{h}}}

\def\bv{{\mathbf{v}}}

\def\b0{{\mathbf{0}}}

\newcommand{\nn}{\nonumber}

\newcounter{probno}

\begin{document}
\title{Simultaneous
  Information and Power Transfer for Broadband Wireless Systems} \author{Kaibin
  Huang and Erik G. Larsson\thanks{
    K. Huang is with the Dept. of Applied Mathematics, the Hong Kong Polytechnic University, Hong Kong,  and
    E. G. Larsson is with the Dept. of Electrical Engineering (ISY), Link\"oping University, Sweden.  Email:
    huangkb@ieee.org, erik.larsson@isy.liu.se.  Updated on \today.  }}

\maketitle

\begin{abstract} 
  Far-field \emph{microwave power transfer} (MPT) will free wireless
  sensors and other mobile devices from the constraints imposed by
  finite battery capacities. Integrating MPT with wireless
  communications to support \emph{simultaneous wireless information and power
    transfer} (SWIPT) allows the same spectrum to be used for dual
  purposes without compromising the quality of service. A novel
  approach is presented in this paper for realizing SWIPT in a
  broadband system where orthogonal frequency division multiplexing
  and transmit beamforming are deployed to create a set of parallel
  sub-channels for SWIPT, which simplifies resource
  allocation. Based on  a proposed reconfigurable mobile
  architecture, different system configurations are considered by
  combining single-user/multi-user systems, downlink/uplink information
  transfer, and variable/fixed coding rates. Optimizing the power
  control for these configurations results in a new class of multi-user
  power-control problems featuring the \emph{circuit-power constraints},
  specifying that the transferred power must be sufficiently large to
  support the operation of the receiver circuitry. Solving these
  problems gives a set of power-control algorithms that exploit
  channel diversity in frequency for simultaneously enhancing the
  throughput and the MPT efficiency.  For the system configurations
  with variable coding rates, the algorithms are variants of
  water-filling that account for the circuit-power constraints. The
  optimal algorithms for those configurations with fixed coding rates
  are shown to sequentially allocate mobiles their required power for
  decoding in ascending order until the entire budgeted power is
  spent. The required power for a mobile is derived as simple
  functions of the minimum signal-to-noise ratio for correct decoding,
  the circuit power and sub-channel gains.
\end{abstract}

\begin{keywords}
Power transmission, cellular networks,  power control, energy harvesting, mobile communication.
\end{keywords}

\section{Introduction}

Microwave power transfer (MPT) refers to wirelessly transmitting
energy from one place to another.  Simultaneous wireless information and power
transfer (SWIPT) refers to using the \emph{same} emitted
electromagnetic (EM) wavefield to transport both  energy that is
harvested at the receiver, and information that is decoded by the
receiver.

In the past decades, much research effort has been directed towards
developing MPT for replacing cables in long-distance power transfer
either terrestrially \cite{Brown:RadioWPTHistory:1984} or from solar satellites to
the earth \cite{Mcspadden:SpaceSolarPowerMicrowaveWPT:2002}. This has led to a series
of breakthroughs in microwave technology including high-power
microwave generators and, more importantly in the current context, the
invention of \emph{rectennas} (rectifying antennas) for efficient
RF-to-DC conversion \cite{Brown:RadioWPTHistory:1984}. This technology has been
applied to the design of helicopters and airplanes powered solely by
microwaves \cite{Schlesak:MicrowavePoweredHiAltitudePlatform:1998}.  Most prior research on MPT focuses on the design of compact and efficient rectennas or
similar energy harvesters
\cite{Brown:RadioWPTHistory:1984, Le:FarFieldRFEnergyHarvesting:2008}. More recently there has been
interest in the powering of low-power devices and even
trickle-recharging of certain personal communications devices. There
is already equipment available that does this \cite{P2110}, by
broadcasting omni-directionally with an RF power of about $1$ W, and
harvesting several mW.  With a massive transmitter array, power could be focused so
that the harvested power is increased hundreds of  times.  The power
levels involved are still small, much smaller than the emitted RF
power by some cell phones (up to 2 W for GSM), so absorption by the
human body does not appear to be a fundamental technological
problem. Moreover, various safety precautions could be applied if
deemed important.

With SWIPT, one and the same wave-field is used to transmit energy and
information.  This has several advantages. First, separate
transmission of power and information by time division is suboptimal
in terms of  efficiently using the available power and bandwidth.
SWIPT, by contrast, may exploit integrated transceiver designs.
Second, with SWIPT, interference to the communication systems can be
kept under control. This is especially important in multi-user systems
with many potential receivers who would suffer from interference.  By contrast, traditional microwave
power transfer (MPT) relies on transmission of a single tone (and its
unintended harmonics), which can interfere with communication
links. Furthermore, MPT does not have any dedicated  spectrum. Hence, as
such, for use in existing bands, it \emph{must} be integrated with
communication solutions.

A key application of SWIPT that we foresee is to provide power to, and
communicate with, sensors for which battery replacement is difficult
or even impossible \cite{PikeReseach:WirelessPower:2012}.  Radio-frequency
identification (RFID) tags are one important example.  RFID is already
a very widely used technology, but its full potential is probably not
fully exploited. A major limitation is the small range of RFID readers
with constrained power. Another limitation is the ability of readers
to correctly resolve different RFID tag returns that arrive at the
receiver superimposed on one another.  Many other applications, for
example, in the chemical process industry, in environmental
monitoring, in oil platforms and pipelines and in surveillance and
national security applications require sensors with extreme
reliability. Often these sensors transmit rather modest amounts of
data, in some applications, only a few bits per hour. Typically the
sensors are hard to access and therefore their batteries require long
lifetimes and very low failure rates.

Making SWIPT work will require integration between multiantenna
transmission, efficient energy harvesting, resource management and
signal processing.  In particular, theory and methods for massive MIMO
\cite{RusekLarssonMarz:ScaleUpMIMO:2012} may become fundamental
enablers for SWIPT. Enabling technology for realizing SWIPT in practice
is the theme of this paper.

\subsection{Prior Related Work}

The concept of SWIPT, in a very basic form, has existed for a long time
in applications like RFID and power-line communications. It was first
studied from an information-theoretic perspective in
\cite{Varshney:TransportInformationEnergy:2008} for a narrow-band
noisy channel, and later in
\cite{GroverSahai:ShannonTeslaWlssInfoPowerTransfer} for a
frequency-selective channel.  These papers characterized the
fundamental trade-off between communication capacity and power
harvested at the receiver.  A similar trade-off was derived for a
multi-user system in \cite{FoulSimeone:TransInfoEnergyMU}. From a
communication theoretic point of view, the novel aspect here is the
new constraint on the minimum received power representing the fixed
circuit power consumption, called the \emph{circuit-power constraint},
which results in the said  fundamental tradeoff.

These aforementioned studies implicitly assumed that the received
energy can be still harvested after passing through an information
decoder, which is infeasible given the current state-of-the-art of
electronic circuits. This motivated the design of practical
SWIPT-enabled receivers that \emph{split} the received microwave signal
from each antenna and feed it to two separate circuits, one for
information decoding and one for energy harvesting
\cite{Zhang:MIMOBCWirelessInfoPowerTransfer,ZhouZhang:WlessInfoPowrTransfer:RateEnergy:2012}.
The corresponding capacity-and-energy tradeoffs are characterized for
the multiple-input-multiple-output (MIMO) channels with perfect
transmitter channel state information (CSIT)
\cite{Zhang:MIMOBCWirelessInfoPowerTransfer,
  ZhouZhang:WlessInfoPowrTransfer:RateEnergy:2012} and further
investigated for the case of imperfect CSIT
\cite{XiangTao:RobustBeamformWIPT:2012}. An additional scenario
considered in \cite{Zhang:MIMOBCWirelessInfoPowerTransfer} is
broadcasting from a base station to two receivers taking turns for
information decoding and energy harvesting, corresponding to
\emph{time-division-information-and-power transfer} (TD-IPT). This
protocol simplifies the receiver design but compromises the
efficiencies of MPT and IT since they cannot operate simultaneously.
The systems considered in the aforementioned prior works share the
common setting that a transmitter draws energy from a reliable source
such as the electric grid and then delivers it to passive devices by
MPT. A different scenario related to distributive networks such as
sensor networks is one where devices exchange energy in addition to
peer-to-peer communication. Transmission strategies are proposed in
\cite{Simeone:InterativeTransferEnergyInfo:2012} for two devices to
exchange information and energy based on TD-IPT over a two-way
channel. The principle of energy sharing is also reflected in a relay
system studied in \cite{GurakanUlukus:EnergyCooperation:2012} where a
source node transfers energy to a relay node in return for its
assistance in transmission. It is shown that jointly managing the
energy queues at these nodes that both harvest energy from external
sources can enhance the end-to-end throughput.

Realizing SWIPT in practice requires not only suitable hardware and
physical-layer algorithms but also the support of an appropriate
network architecture. One such architecture, proposed in
\cite{HuangLau:EnableWPTCellularNetworks:2013}, overlays a traditional
cellular network with additional base stations dedicated for MPT to
mobiles. Based on a stochastic-geometry network model and under a
quality-of-service constraint for the data links, a tradeoff is
derived between the densities of the base stations for MPT and those
for IT, giving insight into the optimal network deployment.

A popular modulation method called \emph{orthogonal frequency division
  multiplexing} (OFDM) divides a broadband channel into decoupled
narrowband sub-channels. OFDM simplifies the channel equalization and
multiple access [facilitating \emph{orthogonal frequency division
  multiple access} (OFDMA)], which has motivated its adoption in
modern communication standards such as 3GPP and WiFi
\cite{GoldsmithBook:WirelessComm:05}. Designing SWIPT based on OFDM not
only retains its existing advantages but also enables  simultaneous
wireless recharging of multiple devices. The current work represents a
first attempt to develop a practical framework for OFDM-based SWIPT
that features a practical mobile architecture and a matching set of
power-control algorithms that exploit frequency diversity to enhance
the efficiency of SWIPT. In parallel with  our initial
results in \cite{HuangLarsson:SIPTBroadbandChannel}, an independent
study on the same topic was reported in
\cite{NgLo:MultiuserOFDMSInfoPowerTransfer}. The practicality of the
SWIPT system proposed in \cite{NgLo:MultiuserOFDMSInfoPowerTransfer}
seems to be limited in several respects. First, the use of a
single-antenna base station for SWIPT leads to isotropic radiation of
the transmission power and hence an extremely low MPT efficiency. This is
the reason that beamforming is the primary technology for practical MPT
solutions \cite{Brown:RadioWPTHistory:1984,
  Mcspadden:SpaceSolarPowerMicrowaveWPT:2002,
  Schlesak:MicrowavePoweredHiAltitudePlatform:1998}. Isotropic MPT
also couples the multi-user MPT links and results in difficult power
control problems \cite{NgLo:MultiuserOFDMSInfoPowerTransfer}. Second,
the design in \cite{NgLo:MultiuserOFDMSInfoPowerTransfer} is based on
the assumption that information decoding causes no loss in harvesting
the total received energy. While this assumption is common (see,
e.g., \cite{Varshney:TransportInformationEnergy:2008,
  GroverSahai:ShannonTeslaWlssInfoPowerTransfer,
  FoulSimeone:TransInfoEnergyMU}), we know of no compelling arguments
for its practicality. Lastly, a sub-optimal TD-IPT protocol instead of
SWIPT is adopted in \cite{NgLo:MultiuserOFDMSInfoPowerTransfer}. These
drawbacks of existing approaches  may be overcome by the
SWIPT framework proposed in this paper.

\subsection{Summary of Contributions and Organization}

This work assumes a noise-limited broadband system where a
multi-antenna base station not only communicates with but also
wirelessly powers the mobile devices.  The broadband channel is
partitioned into orthogonal sub-channels by OFDM and the base station
transmits/receives one data stream per sub-channel. Streams are
encoded with either \emph{variable rates} adapted to the receive
signal-to-noise ratios (SNRs) or \emph{fixed rates} for which
successful decoding requires the receive SNRs to exceed a given
threshold. The constraint and threshold are referred to as the
\emph{minimum-SNR constraint} and the \emph{SNR threshold},
respectively.  Assuming sparse scattering and perfect CSIT, the base
station steers beams for different sub-channels towards associated
mobiles, creating a set of parallel channels for SWIPT. Note that OFDM
alone without beamforming can decouple only the IT links but not the
MPT links. The transmission power for different sub-channels is
controlled subject to a constraint on the total power.  We consider
both a single-user system where the mobile is assigned all
sub-channels and a multi-user system where each mobile is assigned a
single sub-channel based on OFDMA. Two practical scenarios for SWIPT
are considered depending on if IT is in the downlink or the uplink
direction. For SWIPT with downlink IT, the OFDM signal transmitted by
the base station is used both for IT and for MPT.  For SWIPT with
uplink IT, MPT and IT are in the opposite directions where downlink
MPT relies on the transmission of unmodulated tones
\cite{Choi:FullDuplexWireless:2010}, called \emph{power tones}, and
uplink data signals are OFDM modulated.  In this scenario,  the base
station is assumed to support \emph{full-duplex} SWIPT based on the
same principle as proposed in
\cite{Choi:FullDuplexWireless:2010}. More specifically, the antenna
array at the base station is partitioned into two sub-arrays for
transmit beamforming and receive combining and the cross-coupled power
tones in the received uplink signal is perfectly canceled. This is
viable since the base station has perfect knowledge of the phases and
frequencies of the power tones.

A SWIPT-enabled mobile architecture is proposed that can be
reconfigured according to the direction of the IT.  The architecture
consists of \emph{dual} antennas, one information transceiver and one
energy harvester. The harvester \emph{continuously} converts incoming
microwaves to DC power which is used to operate the mobile circuit and
to supply transmission  power for the uplink IT. This is feasible by using
existing energy harvester designs such as those in
\cite{ZhouZhang:WlessInfoPowrTransfer:RateEnergy:2012,HuangLau:EnableWPTCellularNetworks:2013}.
When configured for SWIPT with downlink IT, the outputs of the two
antennas are combined and then split using a power splitter with an
adjustable ratio to yield the inputs of the receiver and harvester,
similarly to the designs in
\cite{Zhang:MIMOBCWirelessInfoPowerTransfer,
  ZhouZhang:WlessInfoPowrTransfer:RateEnergy:2012}. The power
splitting ratio provides a degree-of-freedom for managing the received
power for IT and MPT. When the architecture is reconfigured for SWIPT
with uplink IT, the two antennas are separately attached to the
transmitter and harvester and support full-duplex SWIPT in opposite
directions.  The mobile architecture is assumed to consume fixed
circuit power, following practical models
\cite{MiaoLi:CrossLayerOptimEnergyEffSurvey:2009}.  Based on the
transmission scheme and mobile architecture described earlier,
algorithms for power control at the base station are designed for a
comprehensive set of system configurations combining
single-user/multi-user systems, downlink/uplink IT, and variable/fixed
coding rates. The key features of the proposed algorithms are
summarized in Table~\ref{Tab:Algo}.

The remainder of the paper is organized as follows. The system model
is described in Section~\ref{Section:System}. The SWIPT-enabled mobile
architecture is proposed in Section~\ref{Section:Architecture}. Based
on the architecture, power-control algorithms are designed separately
for the four scenarios combining single-user/multi-user systems and
downlink/uplink IT in
Sections~\ref{Section:SU:DLIT}--\ref{Section:MU:ULIT}. Their
performance is evaluated by simulation in
Section~\ref{Section:Simulation}, followed by concluding remarks in
Section~\ref{Section:Conclusion}.

\renewcommand{\arraystretch}{1.5} 
 \begin{table*}
 \caption{Summary of power control algorithms}
 \centering
\begin{tabular}{|c|c|m{1.5cm}|m{13.6cm}|}
  \hline
  Rate & IT  & \# of Users  & \hspace{3cm} Power-Control Algorithm      \bigstrut \\ 
  \hline
  \multirow{4}[4]{*}{Var.}  &  DL & Single or Multiple & The problem of optimal power control is  non-convex but can be approximated by the convex formulation of the classic multi-channel power control such that circuit power is accounted for in the constraints. This results in policies that are  variants of water-filling. \\ \cline{2-4}
  & \multirow{2}[4]{*}{UL}&Single & 
  \vspace{5pt}\begin{itemize}
  \item[--] \emph{Downlink power control:} The optimal policy is to allocate the maximum power over the strongest
    sub-channel for maximizing the downlink-MPT efficiency.
  \item[--] \emph{Uplink power control:} At the mobile, part of the
    harvested power is used for operating the circuit and the
    remainder is used to maximize the uplink throughput by
    water-filling.  \vspace{-10pt}
\end{itemize}
 \\ \cline{3-4}
&& Multiple& \vspace{5pt}\begin{itemize}
\item[--] \emph{Downlink power control:} The key design technique  is to treat power control as one that injects power into a set of \emph{closed-loop} sub-channels,  where the loss for each sub-channel combines the downlink-and-uplink propagation loss and the circuit-power consumption. Based on this technique, a sub-optimal algorithm  is proposed  that first schedules active mobiles using the criterion of maximum MPT efficiency and then allocates power by water filling with the water level depending on the circuit power. 
\item[--] \emph{Uplink power control:} Each mobile spends  the maximum available power. \vspace{-10pt}
\end{itemize}\\ \cline{3-4}
\hline 
\multirow{4}[4]{*}{Fixed}  &  DL & Single or Multiple & Under the the minimum-SNR constraint, the optimal power-control policy is shown to be one that sequentially allocates power to sub-channels in descending order of  the corresponding channel gains, which is called \emph{greedy channel inversion}.  \\ \cline{2-4}
& \multirow{2}[4]{*}{UL}& Single& 
\vspace{5pt}\begin{itemize}
\item[--] \emph{Downlink power control:} The optimal policy follows that for the variable-rate counterpart. 
\item[--] \emph{Uplink power control:} The optimal policy applies greedy channel inversion over the uplink sub-channels. \vspace{-10pt}
\end{itemize}
\\ \cline{3-4}
&& Multiple & 
\vspace{5pt}
\begin{itemize}
\item[--] \emph{Downlink power control:}  The optimal  policy  performs greedy channel inversion based on the effective gains of the said closed-loop sub-channels, which are derived as closed-form functions of the downlink/uplink sub-channel gains and circuit power. 
\item[--] \emph{Uplink power control:}  Each mobile applies the maximum available power for uplink transmission. \vspace{-10pt}
\end{itemize} 
\\ \cline{3-4}
\hline
\end{tabular}\label{Tab:Algo}\vspace{-10pt}
\end{table*}

\begin{figure}[t]
\begin{center}
\includegraphics[width=8cm]{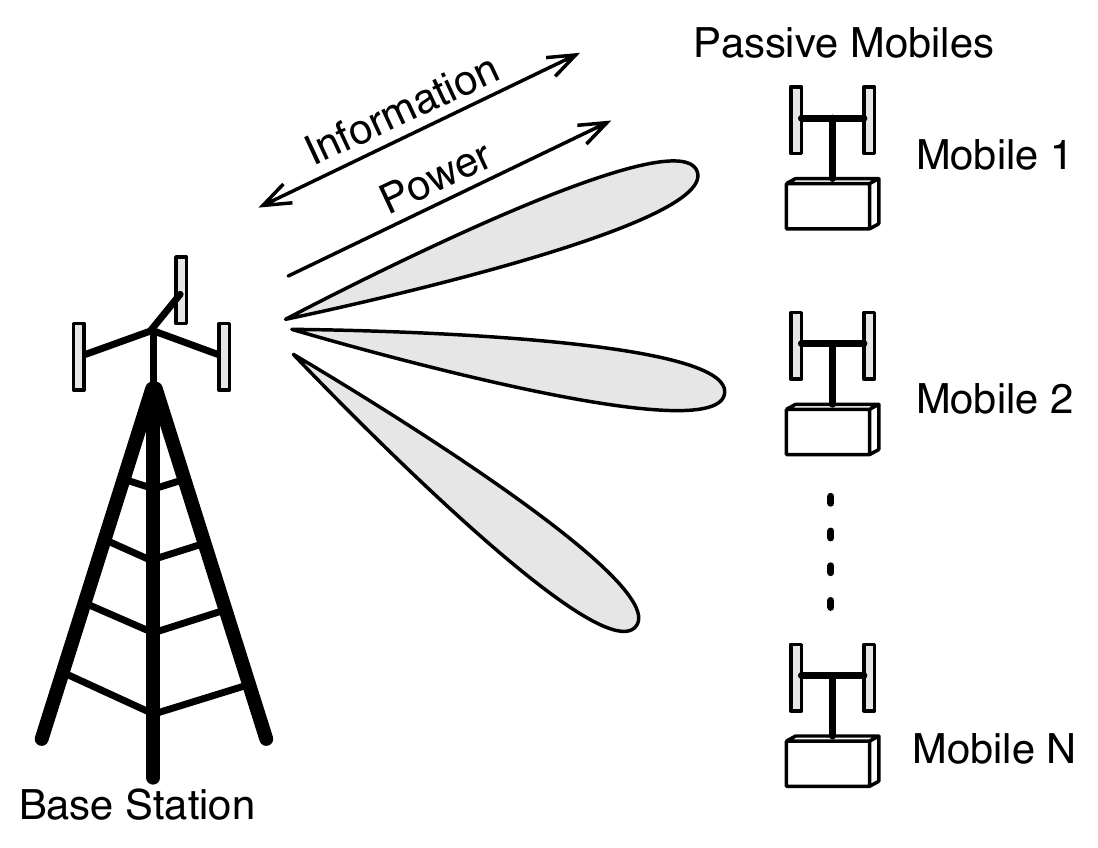}
\end{center}
\caption{SWIPT in a broadband single-cell system where a base station
  serves passive mobiles based on OFDMA. Power is transferred from the
  base station to mobiles. Information transfer can be in either the
  downlink or uplink direction. }
\label{Fig:System}
\end{figure}

\section{System Model}\label{Section:System}

In the single-cell system as illustrated in Fig.~\ref{Fig:System}, a
multi-antenna base station communicates with and supplies power to $N$
mobiles in a sparse-scattering environment. IT takes place in either
the downlink or uplink direction but the MPT is always from the base
station to the mobiles. SWIPT uses a wide spectrum partitioned into $K$
sub-channels. For a single-user system ($N=1$), all sub-channels are
assigned to a single mobile; for a multi-user system, each mobile is
assigned one sub-channel ($N=K$).  Note that the problem formulation
for the case of assigning variable numbers of sub-channels to mobiles
differs from the current one in having more complex circuit-power
constraints, but the solution methods are similar. Ideally, the
sub-channel assignments for the multi-user system should be jointly
optimized with the power control (see, e.g.,
\cite{WonCheETAL:MultOFDMAdapSubc:Oct:99} for traditional OFDMA
systems) but the optimal design for the current scenario seems
intractable due to multi-user-circuit-power constraints. For
tractability, we assume given sub-channel assignments and focus on
the power control. Furthermore, time is slotted and it is assumed for
simplicity that the energy storage of all mobiles are empty at the
beginning of each slot. Consequently, the instantaneous power
harvested by an active mobile is required to meet the circuit-power
constraint. Relaxing the said assumption requires generalizing the
homogeneous circuit-power constraint to heterogeneous ones, which
requires only a straightforward extension of the current results.

\subsection{Coding Rates}

Information streams are transmitted over separate sub-channels and
independently encoded  with either variable
\cite{GoldsmithBook:WirelessComm:05} or fixed coding rates
\cite{Ozarow:InfoTheoCellularMobile:1994}. Given variable coding rates
and perfect CSIT, the rate of a stream is adapted to the receive SNR,
denoted as $\SNR$, and given as $\log_2(1+\SNR)$. Alternatively, the
coding rates can be fixed to $\log_2(1+\theta)$ where the constant
$\theta > 0$ specifies the minimum receive SNR required for correct
decoding.

\subsection{Multi-Antenna Beamforming and Combining} 

We assume an environment with sparse scattering that is necessary for
efficient MPT.  For SWIPT with downlink IT, the antenna array at the
base station is used to reduce the propagation loss by steering beams
towards intended mobiles. Considering an arbitrary slot, let the
vectors $\dot{\bh}_n$ and $\ddot{\bh}_n$ represent particular
realizations of the $n$-th multiple-input-single-output (MISO)
sub-channels from the base station to antenna $1$ and $2$ of the
$n$-th mobile, respectively. Moreover, the transmit beamforming
vector for the $n$-th sub-channel is denoted as $\bff_n$ and computed
by estimating the mobile's direction by training. The beamforming
vectors $\{\bff_n\}$ are assumed given and their designs are outside
the scope of this paper. Then the effective SISO-channel gains
resulting from beamforming can be defined as $\dot{h}_n =
|\bff_n^\dagger\dot{\bh}_n|^2$ and $\ddot{h}_n =
|\bff_n^\dagger\ddot{\bh}_n|^2$.

For SWIPT with uplink IT, the antenna array at the BS is divided into
two sub-arrays. These sub-arrays and the dual antennas at a particular
mobile create a downlink MISO channel and an uplink
single-input-multiple-output (SIMO) channel for supporting the
full-duplex operation of SWIPT. Abusing the notation, let
$\dot{\bh}_n$ denote the $n$-th downlink vector sub-channel and
$\ddot{\bh}_n$ the $n$-th uplink vector sub-channel.  Beamforming and
maximum-ratio combining are applied at corresponding sub-arrays to
enhance the MPT efficiency and the receive SNR of the uplink signal,
respectively. Let $\dot{\bff}_n$ denote the transmit beamforming
vector for the $n$-th downlink sub-channel and $\ddot{\bff}_n$ the
combining vector for the $n$-th uplink sub-channel.  
The effective SISO channels in the opposite directions have the gains defined as
$g'_n = |\dot{\bff}_n^\dagger\dot{\bh}_n|^2$ and $g_n =
|\ddot{\bff}_n^\dagger\ddot{\bh}_n|^2$.

\subsection{Broadband Signals} 

\begin{figure}[t]
\begin{center}
\subfigure[]{\includegraphics[width=8.5cm]{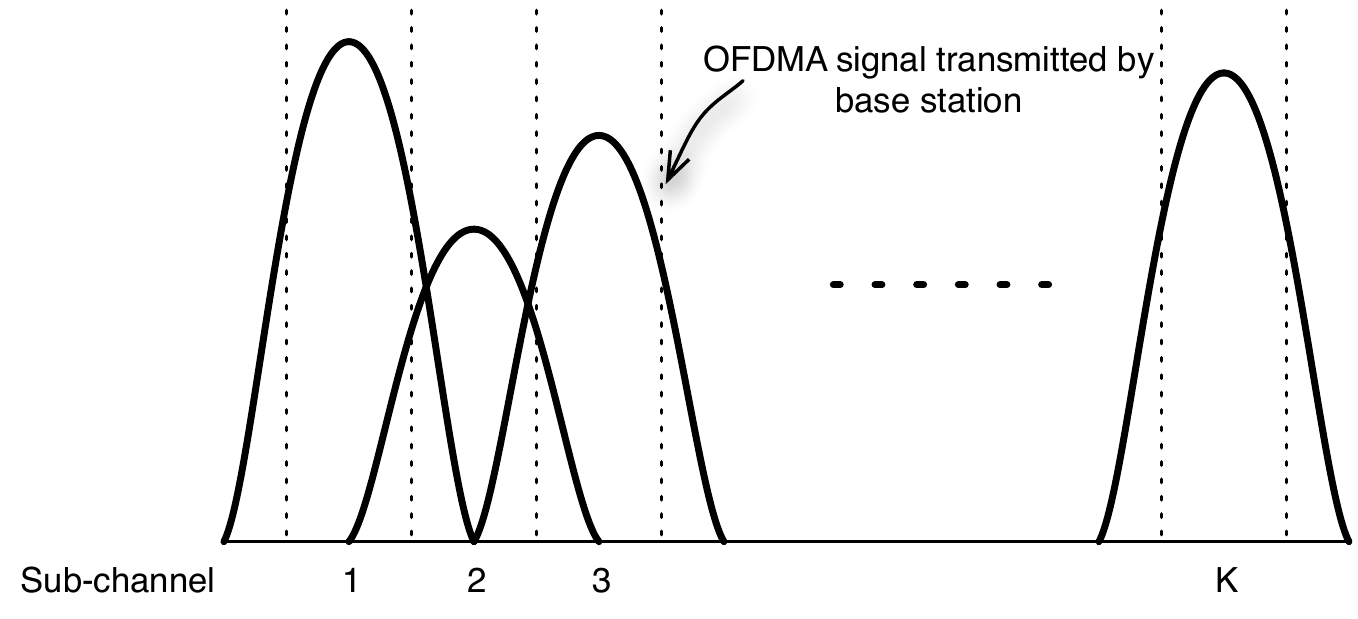}}\\
\subfigure[]{\includegraphics[width=8.5cm]{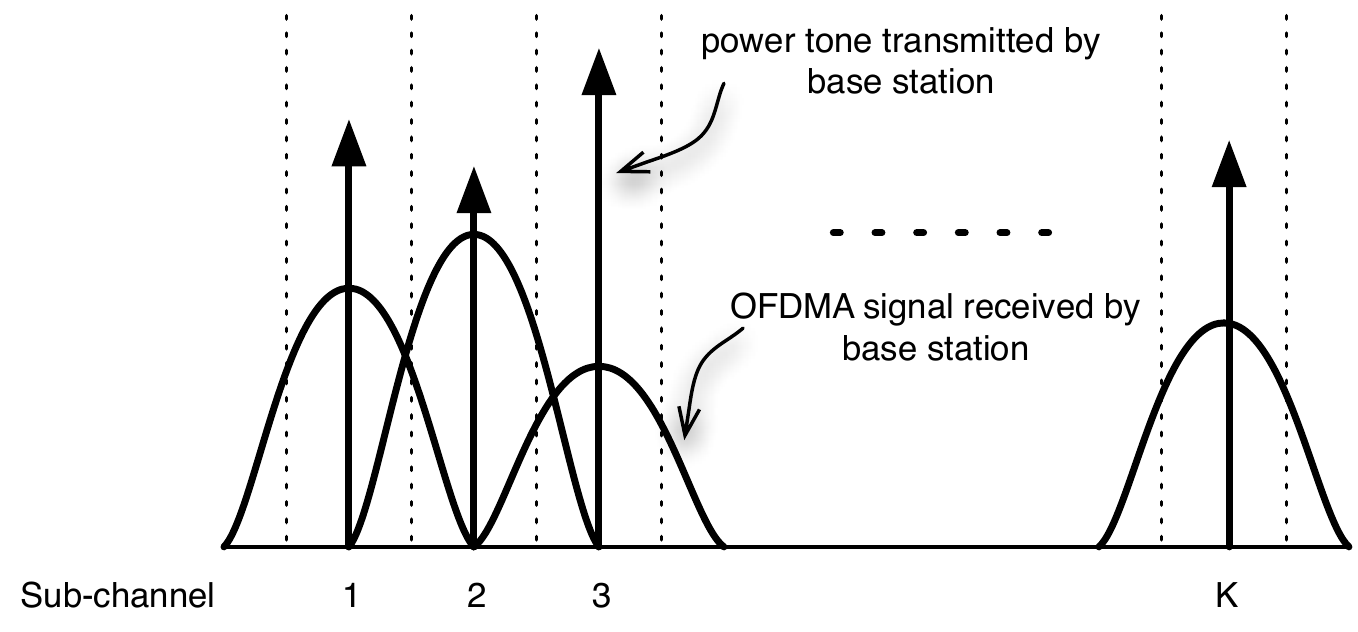}}
\end{center}\vspace{-5pt}
\caption{(a) Spectrum for SWIPT with downlink IT where the downlink signal is OFDM modulated and there is no uplink transmission. (b) Spectrum for SWIPT with uplink IT realized by OFDMA signals transmitted by mobiles while  downlink MPT uses power tones transmitted by the base station.  }
\label{Fig:Spectrum}
\end{figure}

Consider SWIPT with downlink IT. For this scenario, the data-bearing
signal transmitted by the base station is OFDM modulated as
illustrated in Fig.~\ref{Fig:Spectrum}(a).  Due to either  safety
regulations or  limitations of the base-station hardware, the powers
allocated over the sub-channels, denoted as $\{P_n\}$, satisfy a
power constraint:
\begin{equation}\label{Eq:PeakPwr:Const}
\sum\nolimits_{n=1}^K P_n \leq p_t
\end{equation}
where $p_t > 0$ represents the maximum total transmission power. A
mobile extracts information and energy from the same received signal
using the receiver architecture discussed in the next section.

Next, consider SWIPT with uplink IT. As illustrated in
Fig.~\ref{Fig:Spectrum}(b), downlink MPT relies on the transmission of
$K$ power tones at the centers of  the corresponding
sub-channels and their sum power satisfies the power constraint
in \eqref{Eq:PeakPwr:Const}. For a single-user system, the $K$ power
tones are beamed by the base station to a mobile. Besides operating
the circuit, the mobile uses part of the harvested power to enable
uplink IT where the uplink data signal is OFDM modulated as shown in
Fig.~\ref{Fig:Spectrum}(b).  For a multi-user system, the $K$ power
tones are beamed to $K$ corresponding mobiles. The uplink transmission
by the mobiles is based on OFDMA.

\section{SWIPT-Enabled Mobile Architecture}\label{Section:Architecture}\label{Section:Mobile}

\begin{figure}[t]
\centering
\includegraphics[width=9cm]{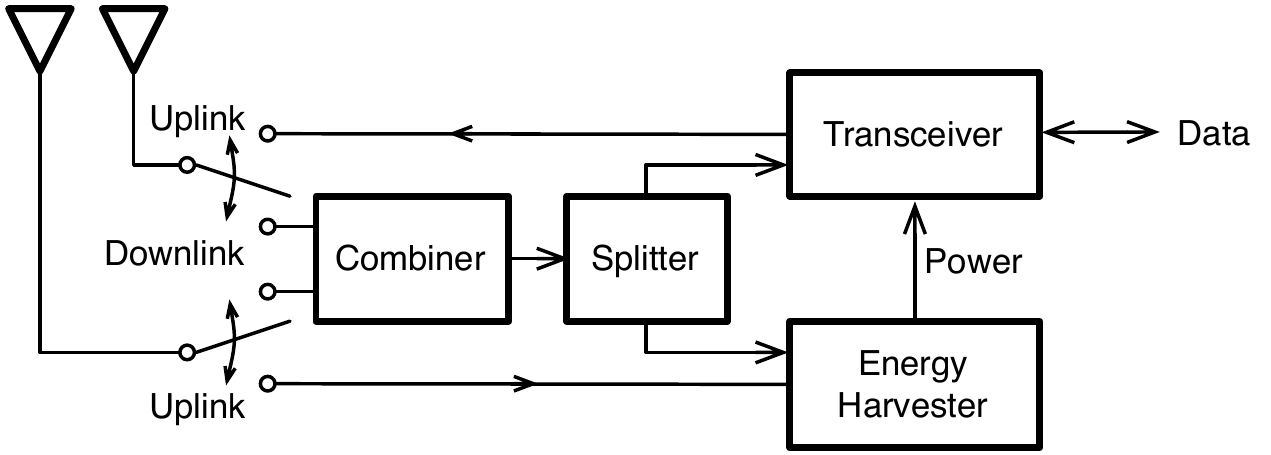}
\caption{Reconfigurable mobile architecture that supports
  dual-mode SWIPT, namely, SWIPT with downlink or uplink IT.  }
\label{Fig:Transceiver}
\end{figure}

In this section, we propose a dual-antenna mobile architecture as
illustrated in Fig.~\ref{Fig:Transceiver} for supporting dual-mode
SWIPT. The architecture comprises a transceiver and an energy
harvester. The transceiver demodulates and decodes received data for
downlink IT or encodes and modulates data for uplink IT.  The energy
harvester converts the input signal into DC power for operating the
circuitry. The architecture can be reconfigured depending on whether
the IT takes place in the uplink or downlink direction.

Consider the architecture configured for downlink IT. The antenna
outputs are then coherently combined to enhance the received signal
power (see Fig.~\ref{Fig:Transceiver}). The combiner output is split
into inputs to the receiver and to the energy harvester
\cite{ZhouZhang:WlessInfoPowrTransfer:RateEnergy:2012}.  To be
specific, the received signal is split using a power splitter that
multiplies the signal with the adjustable factors $\sqrt{\beta}$ and
$\sqrt{1-\beta}$, where $\beta \in [0, 1]$, in order to obtain the
inputs to the receiver and the energy harvester, respectively.
Consequently, the received power is divided into two parts of relative
magnitudes $\beta$ and $(1-\beta)$.  Let $\sigma^2_a$ and $\sigma_b^2$
represent the variances of the noise for a sub-channel, as accumulated
in the path before and after the splitter, respectively. To simplify
notation, we assume that the total noise has unit variance and thus
$\sigma^2_a+ \sigma_b^2=1$. Using these definitions, the receive SNR
for the $n$-th stream can be written as
\cite{Zhang:MIMOBCWirelessInfoPowerTransfer}
\begin{equation}\label{Eq:SNR:DLIT}
\SNR_n = \frac{\beta P_n h_n }{\beta \sigma_a^2 + \sigma_b^2} 
\end{equation}
where $h_n = \dot{h}_n + \ddot{h}_n$ due to the maximum-ratio
combining. Neglecting the small contributions from noise and beam
sidelobes, and assuming lossless RF-to-DC conversion, the harvested
power at a mobile is $(1-\beta) \sum_{n=1}^K P_n h_n$ for a
single-user system and $(1-\beta) P_n h_n$ for a multi-user system
where the mobile is assigned the $n$-th sub-channel.

Next, for the mobile architecture configured for uplink IT, two
antennas are separately attached to the transceiver and energy
harvester to support the  full-duplex operation of the information and power
transfers in the opposite directions (see
Fig.~\ref{Fig:Transceiver}). Under the assumption of unit noise
variance, the receive SNR at the base station for the $n$-th stream is
$\SNR_n = Q_n g_n$ where $Q_n$ represents the uplink-transmission
power allocated to the $n$-th sub-channel. The harvested power at a
mobile is $\sum_{n=1}^K P_n g'_n$ for the single-user system and $P_n
g'_n$ for the multi-user system when the mobile is assigned the $n$-th
sub-channel.

Finally, it is worth mentioning that adding more antennas at a mobile enhances the
received signal power by increasing the total antenna aperture as well
as providing an array gain for the uplink transmission. Nevertheless, spatial
multiplexing is difficult since a typical environment for efficient MPT has line-of-sight and the corresponding channel matrix is practically rank-one. 

\section{Power Control for Single-User SWIPT Systems with Downlink
  IT}\label{Section:SU:DLIT}

\subsection{Single-User Downlink IT with Variable Coding Rates}

\subsubsection{Problem formulation} Given the receive SNR in
\eqref{Eq:SNR:DLIT}, the downlink throughput, denoted as $C_v$, can be
written as
\begin{equation}\label{Eq:Cap:SUDL:Var}
C_v \!=\! \l[\sum_{n=1}^K \log_2\! \l(\! 1 + \frac{\beta P_n h_n}{\beta \sigma^2_a \! +\! \sigma^2_b}\r)\r]I\l(\!(1\!-\!\beta) \sum_{n=1}^K P_n h_n \geq p_c\!\r)
\end{equation}
where the indicator function $I(\mathcal{E})$ gives $1$ if the event
$\mathcal{E}$ occurs and $0$ otherwise. The indicator function in
\eqref{Eq:Cap:SUDL:Var} represents the circuit power constraint.  The
problem of maximizing the throughput in \eqref{Eq:Cap:SUDL:Var} by
power control is formulated as:
\begin{equation}
\text{({\bf P1})} \quad \begin{aligned}
\underset{\beta, \{P_n\}}{\text{max}}\quad &  \sum\nolimits_{n=1}^K \log\l(1 + \frac{\beta P_n h_n}{\beta \sigma^2_a + \sigma^2_b}\r)\\
\text{s.t.} \quad 
& \sum\nolimits_{n=1}^K P_n \leq p_t,\\
& (1-\beta) \sum\nolimits_{n=1}^K P_n h_n \geq p_c, \\
&P_n \geq 0 \ \forall \ n, \\
& \beta \in [0, 1].  
\end{aligned}\nn
\end{equation}

\subsubsection{Solution} P1 is non-convex but can be approximated by
a convex problem as follows. Since, by assumption, $\sigma^2_a+
\sigma_b^2=1$, the rate function in P1 is bounded as
\begin{equation}\label{Eq:Cap:SUDL:Var:Bnds}
\begin{aligned}
\log\l(1 + \beta P_n h_n\r) &\leq \log\l(1 + \frac{\beta P_n h_n}{\beta \sigma^2_a + \sigma^2_b}\r) \\
&\leq \log\l(1 + \frac{\beta P_n h_n}{\sigma^2_b}\r).
\end{aligned}  
\end{equation}
Approximating the objective function in P1 using the lower bound in \eqref{Eq:Cap:SUDL:Var:Bnds} yields 
\begin{equation}
\text{({\bf P1.1})} \quad \begin{aligned}
\underset{\beta, \{P_n\}}{\text{max}}\quad &  \sum\nolimits_{n=1}^K \log\l(1 + \beta P_n h_n\r) \\
\text{s.t.} \quad 
&P_n \geq 0 \ \forall \ n, \\
& \sum\nolimits_{n=1}^K P_n \leq p_t,\\
& (1-\beta) \sum\nolimits_{n=1}^K P_n h_n \geq p_c, \\
& \beta \in [0, 1].  
\end{aligned}\nn
\end{equation}
The alternative approximation using the upper bound in
\eqref{Eq:Cap:SUDL:Var:Bnds} has the same structure as P1.1 and hence
is omitted for brevity. Moreover, both approximations give nearly
optimal power control policies as showed by simulation.  P1.1 is a
convex problem and can be solved numerically by standard algorithms
for convex optimization \cite{BoydBook}.  In the remainder of this
section, we investigate the structure of the power control policy that
solves P1.1.

First, it is necessary to test the feasibility of powering the
receiver given transmission power $p_t$. This requires computing the
limit of the harvested power $p_{\max}$ by solving the following
optimization problem:
\begin{equation}
\text{({\bf P1.2})} \quad \begin{aligned}
\underset{\{P_n\}}{\text{max}}\quad & \sum\nolimits_{n=1}^K P_n h_n\\
\text{s.t.} \quad 
& \sum\nolimits_{n=1}^K P_n \leq p_t,\\
&P_n \geq 0 \ \forall \ n. \\
\end{aligned}\nn
\end{equation}
By inspecting P1.1, it is found that $p_{\max} = p_t\max_{n}h_n$. 
It follows that SWIPT is feasible if and only if 
\begin{equation}\label{Eq:SU:Feasible}
p_t \geq \frac{p_c}{\max_{n}h_n}. 
\end{equation}

Next, given the feasibility condition in \eqref{Eq:SU:Feasible}, fixing $\beta$ in  P1.1 leads to 
\begin{equation}
\text{({\bf P1.3})} \quad \begin{aligned}
\underset{\{P_n\}}{\text{max}}\quad & \sum\nolimits_{n=1}^K \log\l(1 + \beta P_n h_n\r)\\
\text{s.t.} \quad 
& \sum\nolimits_{n=1}^K P_n \leq p_t,\\
& (1-\beta)\sum\nolimits_{n=1}^K P_n h_n \geq p_c,\\ 
&P_n \geq 0 \ \forall \ n. 
\end{aligned}\nn
\end{equation}
P1.3 can be solved using the method of duality and the solution
$\{\tilde{P}^*_n(\beta)\}$ is \cite{BoydBook}
\begin{equation}\label{Eq:Optm:Pwr:SU:DL:Var:a}
\tilde{P}^*_n(\beta) =\l\{\begin{aligned}
& \frac{1}{\lambda^*(\beta) - \mu^*(\beta) (1-\beta) h_n} - \frac{1}{\beta h_n}, &&n \in\mathcal{O}(\beta)\\
&0,  && \text{otherwise} 
\end{aligned}\r.
\end{equation}
where the set $\mathcal{O}(\beta)$ is chosen to ensure
$\{\tilde{P}_n^*(\beta)\}$ being non-negative, and the positive scalars
$\lambda^*(\beta)$ and $\mu^*(\beta)$ are the Lagrange multipliers
solving the dual problem -- the unconstrained minimization of
the following convex function \cite{BoydBook}:
\begin{equation}\label{Eq:Dual}\nn
\begin{aligned}
 \sum\nolimits_{n\in\mathcal{O}(\beta)}\log(\lambda - \mu h_n) &+ \lambda\l(p_t + \frac{1}{\beta}\sum\nolimits_{n\in\mathcal{O}(\beta)}\frac{1}{h_n}\r) - \\
 &\mu\l(\frac{p_c}{1-\beta} + \frac{|\mathcal{O}(\beta)|}{\beta}\r). 
\end{aligned}
\end{equation}
Note that the index set $\mathcal{O}(\beta)$ in
\eqref{Eq:Optm:Pwr:SU:DL:Var:a} can be obtained by repetitively removing
from $\mathcal{O}(\beta)$ the index of the mobile corresponding to the
smallest negative element of $\{\tilde{P}^*_n(\beta)\}$ and then
recomputing $\lambda^*(\beta), \mu^*(\beta)$ and
$\{\tilde{P}_n^*(\beta)\}$ till the set $\{\tilde{P}_n^*(\beta)\}$
contains only nonnegative elements. It follows from
\eqref{Eq:Optm:Pwr:SU:DL:Var:a} that there exists a $\beta^*\in[0, 1]$
such that the solution to P1.3, denoted as $\{\tilde{P}^*_n\}$, can
be written as
\begin{equation}\label{Eq:Optm:Pwr:SU:DL:Var}
\tilde{P}^*_n = \l\{\begin{aligned}
& \frac{1}{\lambda^* - \mu^* (1-\beta^*) h_n} - \frac{1}{\beta^* h_n}, &&n \in\mathcal{O}\\
&0,  && \text{otherwise}
\end{aligned}\r.
\end{equation}
where $\lambda^* = \lambda^*(\beta^*)$, $\mu^* = \mu^*(\beta^*)$ and
$\mathcal{O} = \mathcal{O}(\beta^*)$.  As a result, the optimal
power-control policy $\{P^*_n\}$ for the current case can be
approximated as $P^*_n \approx \tilde{P}^*_n$ for all $n$.  Simulation
shows that such an approximation yields a throughput very close to the
maximum possible. The power allocation in
\eqref{Eq:Optm:Pwr:SU:DL:Var} can be interpreted as water-filling in
frequency with a water level that decreases with an increasing
sub-channel gain or vice versa. This agrees with the intuition that
less transmission power is required for turning on a receiver if the
MPT loss is smaller. In contrast, the classic water-filling has a
constant water level.

\subsection{Single-User Downlink IT with Fixed Coding Rates}

\subsubsection{Problem formulation} The downlink throughput, denoted
as $C_f$, is proportional to the number of successfully transmitted
streams. Specifically, using the receive SNR in \eqref{Eq:SNR:DLIT},
$C_f$ is written as
\begin{equation}
\begin{aligned}
C_f = \log_2(1+\theta) &I\l((1-\beta) \sum\nolimits_{n=1}^KP_n h_n \geq p_c\r)\times\\
&\sum\nolimits_{n=1}^KI\l(\frac{\beta P_nh_n }{\beta\sigma_a^2 + \sigma_b^2}\geq \theta \r) 
\end{aligned}\label{Eq:Rate:DL:SU:Share}
\end{equation}
where the first indicator function represents the circuit-power
constraint and the sum gives the number of correctly decoded
streams. The problem of maximizing $C_f$ by power control is hence
formulated as:
\begin{equation}
\text{({\bf P2})} \qquad \begin{aligned}
\underset{\beta, \{P_n\}}{\text{max}}\quad & I\l((1-\beta) \sum\nolimits_{n=1}^KP_n h_n \geq p_c\r)\times\\
&\quad \sum\nolimits_{n=1}^KI\l(\frac{\beta P_nh_n }{\beta\sigma_a^2 + \sigma_b^2}\geq \theta \r)\\
\text{s.t.} \quad 
& \sum\nolimits_{n=1}^K P_n \leq p_t,\\
&P_n \geq 0 \ \forall \ n. 
\end{aligned}\nn
\end{equation}

\subsubsection{Solution} Solving P2 is equivalent to finding the
maximum number of successfully transmitted streams, denoted as $k^*$
and derived as follows. First, rearrange the sequence of channel gains 
$\{h_n\}$ in  descending order and denote the result as
$\bar{h}_1, \bar{h}_n, \cdots, \bar{h}_K$. The corresponding
transmission powers are $\bar{P}_1, \bar{P}_n, \cdots, \bar{P}_K$. This
reordering can be represented by the permutation matrix
$\boldsymbol{\Pi}_h$ such that
\begin{equation}
[\bar{h}_1, \bar{h}_n, \cdots, \bar{h}_K]^T = \boldsymbol{\Pi}_h\times [h_1, h_n, \cdots, h_K]^T
\end{equation}
where the superscript $T$ denotes the matrix transposition.  Assume
that $k$ streams are successfully transmitted. Let $\bar{P}_n^*(k)$
represent the power needed to successfully transmit the $n$-th stream
such that the total power is minimized. To this end, it is desirable
to transmit the streams over $k$ sub-channels with the largest channel
gains, namely, $\{\bar{h}_1, \bar{h}_2, \cdots,
\bar{h}_k\}$. Therefore, considering the minimum-SNR constraint,
$\{\bar{P}^*(k)\}$ solves the following optimization problem:
\begin{equation}
\text{({\bf P2.1})} \quad \begin{aligned}
\underset{\{\bar{P}_n\}}{\text{min}}\quad & \sum\nolimits_{n=1}^k \bar{P}_n\\
\text{s.t.} \quad 
& \bar{P}_n \geq \frac{\theta(\beta\sigma^2_a + \sigma^2_b)}{\beta \bar{h}_n}\ \forall \ 1\leq n\leq k,\\
& \sum\nolimits_{n=1}^k \bar{P}_n \bar{h}_n \geq \frac{p_c}{1-\beta },\\
&\beta \in [0, 1]. 
\end{aligned}\nonumber
\end{equation}
Replacing the the inequality constraints in P2.1 results in an
optimization problem with a smaller domain:
\begin{equation}
\text{({\bf P2.2})} \quad \begin{aligned}
\underset{\{\bar{P}_n\}}{\text{min}}\quad & \sum\nolimits_{n=1}^k \bar{P}_n\\
\text{s.t.} \quad 
& \bar{P}_n = \frac{\theta(\beta\sigma^2_a + \sigma^2_b)}{\beta \bar{h}_n}\ \forall \ 1\leq n\leq k, \\
& \sum\nolimits_{n=1}^k \bar{P}_n \bar{h}_n = \frac{p_c}{1-\beta }, \\
&\beta \in [0, 1]. 
\end{aligned}
\end{equation}
Comparing P2.1 and P2.2 reveals that if the domain of P2.2 is
nonempty, the solution to P2.2 must also solve P2.1. The existence of
a solution for P2.2 can be tested by solving the system of linear
equations from the equality constraints. As a result, $\beta$
satisfies the following quadratic equation:
\begin{equation}\label{Eq:Beta:Eq}
\beta^2  - c(k) \beta - d = 0
\end{equation}
where the coefficients $c(k)$ and $d$ are
\begin{equation}\label{Eq:Beta:Eq:Coeff}
c(k) = 1 - \frac{\sigma_b^2}{\sigma_a^2} - \frac{p_c}{k \theta \sigma_a^2} , \qquad d = \frac{\sigma_b^2}{\sigma_a^2}. 
\end{equation}
Solving the equation in \eqref{Eq:Beta:Eq} and choosing the positive
root give the optimal value of $\beta$ for a given $k$, denoted as
$\beta^*(k)$: 
\begin{equation}\label{Eq:Beta:Op}
\beta^*(k) = \frac{c(k) + \sqrt{c^2(k) + 4d}}{2}. 
\end{equation}
Since the quadratic function on the left hand side of
\eqref{Eq:Beta:Eq} is negative for $\beta = 0$ and positive for $\beta
= 1$, $\beta^*(k)$ lies in the range $[0, 1]$ and hence is a valid
value for the splitting ratio. This confirms the existence of a unique
solution for P2.2 (equivalently P2.1) that follows from the equality
constraints in P2.2 as
\begin{equation}\label{Eq:Power:Op:SU:UL:FixCode}
\bar{P}_n^*(k) = \frac{\theta\l[\beta^*(k)\sigma^2_a + \sigma^2_b\r]}{\beta^*(k) \bar{h}_n} 
\end{equation}
and the minimum transmission power for supporting $k$ streams is hence
$\sum_{n=1}^k\bar{P}_n^*(k)$. In other words, the optimal policy
performs greedy channel inversion.

We can now solve P2 by obtaining $k^*$ as the maximum value of
$k$ under the power constraint from \eqref{Eq:PeakPwr:Const},
which involves a simple search. To be specific
\begin{equation}\label{Eq:k:OP}
k^* = \max_k k,\qquad \text{subject to}\ \sum\nolimits_{n=1}^k\bar{P}_n^*(k)\leq p_t 
\end{equation}
with $\bar{P}_n^*(k)$ given in \eqref{Eq:Power:Op:SU:UL:FixCode}. Note
that $k^* = 0$ if $\bar{P}_n^*(1) > p_t$ for which it is infeasible to
transmit any stream. It follows from \eqref{Eq:k:OP} that the solution
to P2, $\{P^*_n\}$ is given as
\begin{equation}\label{Eq:Power:Op:SU:UL:FixCode:a}
\begin{aligned}
&[P^*_1, P^*_2, \cdots, P^*_K]^T = \boldsymbol{\Pi}_h^{-1}\times \\
&\qquad [\bar{P}_1^*(k^*), \bar{P}_2^*(k^*), \cdots, \bar{P}_{k^*}^*(k^*), 0, \cdots, 0]^T. 
\end{aligned}
\end{equation}
The main results of this section are summarized in the following
proposition.

\begin{proposition}\emph{For the single-user SWIPT system with downlink
    IT and fixed coding rates, the optimal power-control policy
    $\{P^*_n\}$ is given in \eqref{Eq:Power:Op:SU:UL:FixCode:a} and
    the corresponding power-splitting ratio is $\beta^*(k^*)$ with
    $\beta^*(k)$ and $k^*$ given in \eqref{Eq:Beta:Op} and
    \eqref{Eq:k:OP}, respectively.}
\end{proposition}

\section{Power Control for Single-User SWIPT Systems with Uplink IT}\label{Section:SU:ULIT}

\subsection{Single-User Uplink IT with Variable Coding Rates}

\subsubsection{Problem formulation} Uplink transmission is feasible
provided that the harvested power exceeds the circuit power: $
\sum_{n=1}^K P_n g'_n\geq p_t$. Under this condition, the total uplink
transmission power is $(\sum_{n=1}^K P_n g'_n - p_c)$ that is
allocated over sub-channels for maximizing the uplink throughput. In other
words, the throughput for the current case can be written as
\begin{align}
\!\!\!\!\! R_v &\!=\! \l[\sum\nolimits_{n=1}^K \log_2\l(1 + Q_n g_n\r)\r]I\l(\sum\nolimits_{n=1}^K P_n g'_n \geq p_c\r) \label{Eq:Rate:UL:SU}
\end{align}
where $\{P_n\}$ satisfies the power constraint in
\eqref{Eq:PeakPwr:Const} and $\{Q_n\}$ represents uplink power control
subject to: 
\begin{equation}
\sum\nolimits_{n=1}^KQ_n \leq \sum\nolimits_{n=1}^K P_n g'_n - p_c. \label{Eq:ULPwr:Const}
\end{equation}
Using \eqref{Eq:Rate:UL:SU}, the problem of maximizing the uplink throughput is formulated as 
\begin{equation}
\text{({\bf P3})} \quad \begin{aligned}
\underset{\{Q_n\}}{\text{max}}\quad & \sum\nolimits_{n=1}^K \log\l(1 + Q_n g_n\r)\\
\text{s.t.} \quad 
&\sum\nolimits_{n=1}^K Q_n \leq \sum\nolimits_{n=1}^KP_ng_n' - p_c, \\
&Q_n \geq 0 \ \forall \ n, \\
& \sum\nolimits_{n=1}^K P_n g_n' \geq p_c,\\
& \sum\nolimits_{n=1}^K P_n \leq p_t. 
\end{aligned}\nn
\end{equation}

\subsubsection{Solution}
By inspecting P3, the optimization problem can be decomposed into two
sub-problems:
\begin{equation}
\text{({\bf P3.1})} \quad \begin{aligned}
\underset{\{P_n\}}{\text{max}}\quad & \sum\nolimits_{n=1}^K P_n g_n'\\
\text{s.t.} \quad 
& \sum\nolimits_{n=1}^K P_n \leq p_t,\\
&P_n \geq 0 \ \forall \ n
\end{aligned}\nn
\end{equation}
and 
\begin{equation}
\text{({\bf P3.2})} \quad \begin{aligned}
\underset{\{Q_n\}}{\text{max}}\quad & \sum\nolimits_{n=1}^K \log\l(1 + Q_n g_n\r)\\
\text{s.t.} \quad 
&\sum\nolimits_{n=1}^K Q_n \leq \sum\nolimits_{n=1}^KP_n^*g_n' - p_c, \\
&Q_n \geq 0 \ \forall \ n, \\
& \sum\nolimits_{n=1}^K P_n^* g_n' \geq p_c
\end{aligned}\nn
\end{equation}
where $\{P^*_n\}$ solves P3.1. The two problems have different
objectives: That of P3.1 is to maximize the downlink transferred power and
that of P3.2 is to maximize the uplink throughput. P3.1 is similar to P1.2 and
it is straightforward to show that
\begin{equation}\label{MaxMPT:SU:VarRate}
P^*_n = \l\{
\begin{aligned}
&p_t, && g'_n =  \max_k g'_k,\\
&0, && \text{otherwise}\\
\end{aligned}
\r.
\end{equation}
and that  the transferred power is $p_t \max_n g'_n$. 
It follows that the feasibility condition for the uplink transmission is 
\begin{equation}\label{Eq:Feasible:UL:SU}
p_t  \geq \frac{p_c}{\max_n g'_n} 
\end{equation}
which is similar to that in \eqref{Eq:SU:Feasible}.  Under this
condition and given \eqref{MaxMPT:SU:VarRate}, P3.2 reduces to the
classic multi-channel power control problem with the water-filling
solution $\{Q^*\}$ given by
\begin{equation}\label{Eq:Pwr:SU:DLIT:Var}
Q^*_n = \eta - \frac{1}{g_n}, \qquad n \in \mathcal{A}
\end{equation}
where the set $\mathcal{A}$ contains the indices of the uplink sub-channels assigned
nonzero power, and $\eta$ is the water level chosen such that
$\sum_{n\in \mathcal{A}} Q^*_n = p_t \max_n g'_n - p_c$. The solution
to P3 is summarized in the following proposition.
\begin{proposition}\label{Prop:SU:DLIT:Var}\emph{Consider the
    single-user SWIPT systems with uplink IT and variable coding rates.
\begin{enumerate}
\item The optimal power-control policy at the base station is to
  maximize the MPT efficiency by transferring the maximum power $p_t$
  over a single tone in the downlink sub-channel with the maximum
  effective channel gain, resulting in the transferred power equal to
  $p_t\max_ng'_n$.
\item Uplink transmission is feasible if and only if the condition in
  \eqref{Eq:Feasible:UL:SU} holds. Under this condition, the optimal
  power-control policy for the uplink transmission distributes the
  total power $(p_t \max_n g'_n - p_c)$ over the sub-channels according
  to the water-filling in \eqref{Eq:Pwr:SU:DLIT:Var}.
\end{enumerate}
}
\end{proposition}

 \subsection{Single-User Uplink IT with Fixed Coding Rates}
 
 \subsubsection{Problem formulation} 
 Under the minimum-SNR and the circuit constraints, the uplink
 throughput is given as
\begin{equation}
\!\! R_f = \log_2(1+\theta) I\l(\sum_{n=1}^KP_n g'_n \geq p_c\r)\sum_{n=1}^KI\l(Q_ng_n\geq \theta \r) \label{Eq:Rate:UL:SU:Share}
\end{equation}
where the uplink transmission power $\{Q_n\}$ satisfies the same
constraint as in \eqref{Eq:ULPwr:Const} for the case of
variable coding rates.  The problem of maximizing the throughput
follows from \eqref{Eq:Rate:UL:SU:Share} as:
 \begin{equation}
\text{({\bf P4})} \quad \begin{aligned}
\underset{\{P_n\}}{\text{max}}\quad & I\l(\sum\nolimits_{n=1}^KP_n g'_n \geq p_c\r)\sum\nolimits_{n=1}^KI\l(Q_ng_n\geq \theta \r)\\
\text{s.t.} \quad 
& \sum\nolimits_{n=1}^K P_n \leq p_t,\\
&P_n \geq 0 \ \forall \ n. 
\end{aligned}\nn
\end{equation}

 \subsubsection{Solution}
 Similar to P3, P4 can be decomposed into two sub-problems. The first
 sub-problem maximizes the transferred power in the downlink  and is identical
 to P3.1. It follows that uplink transmission is feasible if and only
 if the condition in \eqref{Eq:Feasible:UL:SU} is satisfied, namely
 that $p_t \max_k g'_k \geq p_c$. Under this condition, the other
 sub-problem is to maximize the uplink throughput, more exactly:
\begin{equation}
\text{({\bf P4.1})} \quad \begin{aligned}
\underset{\{Q_n\}}{\text{max}}\quad & \sum\nolimits_{n=1}^KI\l(Q_ng_n\geq \theta \r)\\
\text{s.t.} \quad 
&\sum\nolimits_{n=1}^K Q_n \leq p_t \max_k g'_k - p_c,\\
&Q_n \geq 0 \ \forall \ n.  
\end{aligned}\nn
\end{equation}
It can be observed from P4.1 that the optimal power allocation should
be again based on greedy channel inversion. Specifically, the optimal
policy attempts to meet the minimum-SNR constraints of the streams
following the descending order of their corresponding sub-channel
gains $\{g_n\}$. To state the policy mathematically, let the
sequence $\bar{g}_1, \bar{g}_2, \cdots, \bar{g}_K$ represent the values of $\{g_n\}$ 
sorted in  descending order. Let
$\boldsymbol{\Pi}_g$ represent the permutation matrix such that
\[
 [\bar{g}_1, \bar{g}_2, \cdots, \bar{g}_K]^T = \boldsymbol{\Pi}_g\times  [g_1, g_2, \cdots, g_K]^T. 
\]
Following the earlier discussion, the power allocated to the
sub-channels with gains $\{\bar{g}_n\}$, denoted as $\{\bar{Q}^*_n\}$,
is given as
\begin{equation}\label{Eq:Pwr:MUDLVarC:Order}
\bar{Q}^*_n = \l\{
\begin{aligned}
&\frac{\theta}{\bar{g}_n}, && 1\leq n \leq k^*\\
&0, && \text{otherwise} 
\end{aligned}
\r.
\end{equation}
where $k^*$,  $1\leq k^*\leq K$, is the maximum number of uplink
streams under the uplink-power constraint obtained from the first 
constraint in P4.1 as
\begin{equation}
\sum\nolimits_{n=1}^{k^*}\bar{Q}^*_n \leq p_t \max_k g'_k  - p_c. \label{Eq:ULPwr:Const:a}
\end{equation}
The solution to P4 is summarized in the following proposition. 

\begin{proposition}\emph{Consider the single-user SWIPT system with uplink  IT and fixed coding rates.  
\begin{enumerate}
\item The optimal power-control policy at the base station is
  identical to that in Proposition~\ref{Prop:SU:DLIT:Var}.
\item Uplink IT is feasible if and only if $p_t\max_ng'_n > p_c$.
  Under this condition, the optimal power-control policy for uplink
  transmission is given as
\begin{equation}\nn
[Q^*_1, Q^*_2, \cdots, Q^*_K]^T = \boldsymbol{\Pi}_g^{-1}\times  [\bar{Q}^*_1, \bar{Q}^*_2, \cdots, \bar{Q}^*_K]^T 
\end{equation}
with $\bar{Q}^*_n$ in \eqref{Eq:Pwr:MUDLVarC:Order}. 
\end{enumerate}
}
\end{proposition}

\section{Power Control for Multi-User SWIPT Systems with Downlink
  IT}\label{Section:MU:DLIT}

\subsection{Multi-User Downlink IT with Variable Coding
  Rates}\label{Section:DL:MU:VarCode}

\subsubsection{Problem formulation} Using the receive SNR in
\eqref{Eq:SNR:DLIT}, the sum throughput is obtained as
\begin{equation}
\tilde{C}_v = \sum_{n=1}^K \log_2\l(1 + \frac{\beta_n P_n h_n }{\beta_n \sigma^2_a + \sigma^2_b}\r)I((1-\beta_n) P_n h_n \geq p_c). \label{Eq:Rate:DL:MU}
\end{equation}
In contrast to the single-user counterpart in
\eqref{Eq:Cap:SUDL:Var} having a single circuit-power constraint, the
sum throughput in \eqref{Eq:Rate:DL:MU} contains multi-user
circuit-power constraints. The corresponding power-control problem is
formulated as follows:
\begin{equation}
\text{({\bf P5})}  \begin{aligned}
\underset{\{\beta_n, P_n\}}{\text{max}} \ \ & \sum_{n=1}^K \log\l(\! 1 \! +\! \frac{\beta_n P_n h_n }{\beta_n \sigma^2_a + \sigma^2_b}\!\r)\! I((1\! -\! \beta_n) P_n h_n \geq p_c)\\
\text{s.t.} \ \
& \sum\nolimits_{n=1}^K P_n \leq p_t, \\
&P_n \in\{0\}\cup \l[p_c/h_n, \infty\r)\quad \forall \ n,  \\
& \beta_n \in [0, 1]\quad \forall \ n.
\end{aligned}\nn
\end{equation}

\subsubsection{Solution} Like P1, P5 is non-convex but can be
approximated by a convex problem by replacing the objective function
by either the lower or upper bounds in
\eqref{Eq:Cap:SUDL:Var:Bnds}. Both approximating problems have the
same structure and yield practically the same solutions as P5, as shown
by simulation. For brevity, we consider only the
approximation of P5 using the lower bound in
\eqref{Eq:Cap:SUDL:Var:Bnds} and hence solving the following problem:
\begin{equation}
\text{({\bf P5.1})} \begin{aligned}
\underset{\{\beta_n, P_n\}}{\text{max}}\ \ & \sum_{n=1}^K \log\l(1 +\beta_n P_n h_n\r)I((1-\beta_n) P_n h_n \geq p_c)\\
\text{s.t.} \ \
& \sum\nolimits_{n=1}^K P_n \leq p_t,\\
&P_n \in\{0\}\cup \l[p_c/h_n, \infty\r)\quad \forall \ n,  \\
& \beta_n \in [0, 1]\quad \forall \ n.
\end{aligned}\nn
\end{equation}
It can be observed from P5.1 that if $P_n h_n \geq p_c$, it is optimal
to choose $\beta_n$ such that the input power to the energy harvester
is $p_c$ since additional power contributes no throughput gain,
corresponding to $\beta_n = 1 - p_c/(P_n h_n)$. Consequently, P5.1 can be
rewritten as
\begin{equation}
\text{({\bf P5.2})} \quad \begin{aligned}
\underset{\{P_n\}}{\text{max}} \quad & \sum\nolimits_{n=1}^K \log\l(1 - p_c + P_n h_n\r)I(P_n h_n \geq p_c)\\
\text{s.t.} \quad 
& \sum\nolimits_{n=1}^K P_n \leq p_t, \\
&P_n \in\{0\}\cup \l[p_c/h_n, \infty\r)\quad \forall \ n. 
\end{aligned}\nn
\end{equation}
Let $\mathcal{B}$ denote the indices of the mobiles that meet their
circuit-power constraints using the power allocation in the solution of
P5. Given $\mathcal{B}$ and defining $T_n = P_n - p_c/h_n$, P5.2 can
be simplified as
\begin{equation}
\text{({\bf P5.3})} \quad \begin{aligned}
\underset{ \{T_n\}}{\text{max}}\quad & \sum\nolimits_{n\in \mathcal{B}} \log\l(1 + T_n h_n\r)\\
\text{s.t.} \quad 
& \sum\nolimits_{n\in \mathcal{B}} T_n \leq p_t - p_c\sum\nolimits_{n\in\mathcal{B}} \frac{1}{h_n},\\
&T_n\geq 0 \quad \forall \ n\in\mathcal{B}.  
\end{aligned}\nn
\end{equation}
As the values of $\{h_n\mid n \in \mathcal{B}\}$ increase, the
objective function in P5.3 increases and the first constraint is
relaxed. It follows that with $L = |\mathcal{B}|$, P5.3 is equivalent
to
\begin{equation}
\text{({\bf P5.4})} \quad \begin{aligned}
\underset{\{\bar{T}_n\}}{\text{max}} \quad & \sum\nolimits_{n=1}^L \log\l(1 + \bar{T}_n \bar{h}_n\r)\\
\text{s.t.} \quad 
& \sum\nolimits_{n=1}^L \bar{T}_n \leq p_t - p_c\sum\nolimits_{n=1}^L \frac{1}{\bar{h}_n}\\
&\bar{T}_n\geq 0 \quad \forall \ 1\leq n\leq L  
\end{aligned}\nn
\end{equation}
where $[\bar{T}_1, \bar{T}_2, \cdots, \bar{T}_K]^T =
\boldsymbol{\Pi}_h [T_1, T_2, \cdots, T_K]^T$. The form of P5.4 is
similar to that of the traditional  multi-channel power control problem
with the key difference that the maximum of $\sum_{n=1}^L \bar{T}_n$
increases with decreasing $L$. The reason is that reducing the number
of streams decreases the total circuit-power consumption of the system
and thereby allows more power to be used for IT. Given $L$, combining
the traditional water-filling method and the constant $\bar{T}^*_n = -
p_c/h_n$ if $n > L$ yields the solution to P5.4 as follows:
\begin{equation}\label{Eq:Power:MU:DL:Var}
\bar{T}^*_n = \l\{
\begin{aligned}
&\frac{p_t + (1 - p_c)\sum_{n=1}^{L}\frac{1}{\bar{h}_n}}{L} - \frac{1}{\bar{h}_n}, && 1\leq n \leq L, \\
&-\frac{p_c}{\bar{h}_n}, && \text{otherwise}.
\end{aligned}
\r. 
\end{equation}
The corresponding sum throughput is 
\begin{equation}
\tilde{C}^*_v(L) = \sum_{n=1}^{L}\log_2(\bar{h}_n) + \log_2\l(\frac{p_t + (1 - p_c)\sum_{n=1}^{L}\frac{1}{h_n}}{L}\r). \nn
\end{equation}

Next, the number of streams $L$ is determined by a simple
search. According to the traditional water-filling method, $L$ is
chosen as $L= \ell_{\max}$ where $\ell_{\max}$ with $1\leq
\ell_{\max}\leq K$ is the largest integer such that
$\l\{\bar{T}^*_n\mid 1\leq n \leq \ell_{\max}\r\}$ are positive. It is
important to note that the traditional choice may not be optimal due
to the aforementioned difference between the traditional method and
P5.4. In other words, reducing the number of streams below
$\ell_{\max}$ may result in a throughput gain. The optimal value of
$L$, however, has no closed-form solution but can be obtained by a
simple search over the range from $1$ to $\ell_{\max}$. To be
specific, the value of $L$ that maximizes the sum throughput is given
as
\begin{equation}\label{Eq:Comp:L}
L^* = \arg\max_{1\leq \ell \leq \ell_{\max}}\tilde{C}^*_v(\ell).
\end{equation}
The above results are summarized in the following lemma.
\begin{lemma}\label{Lem:ConvApprox}\emph{The solution for P5.1, denoted as $\{\tilde{P}^*_n\}$, is obtained from  $\{\bar{T}^*_n\}$ in \eqref{Eq:Power:MU:DL:Var} as 
\begin{equation}
\begin{aligned}
\l[\tilde{P}^*_1, \tilde{P}^*_2, \cdots, \tilde{P}^*_K\r]^T = &\boldsymbol{\Pi}_h^{-1} \Bigg\{\l[\bar{T}^*_1, \bar{T}^*_2, \cdots, \bar{T}^*_K\r]^T  + \\
&\quad \l.\l[\frac{p_c}{\bar{h}_1}, \frac{p_c}{\bar{h}_2}, \cdots, \frac{p_c}{\bar{h}_K}\r]^T \r\}  
\end{aligned}
\nn
\end{equation}
with $\{\bar{T}^*_n\}$  in \eqref{Eq:Power:MU:DL:Var} and the number of active mobiles optimized as in \eqref{Eq:Comp:L}. 
}
\end{lemma}
Since P5.1 is a convex approximation of P5, the solution $\{P^*_n\}$
or equivalently the optimal power-control policy for the current case
can be approximated as $P^*_n \approx \tilde{P}^*_n$ for all $n$,
which is shown by simulation to be close-to-optimal.

\subsection{Multi-User Downlink IT with Fixed Coding Rates}

\subsubsection{Problem formulation} Using the receive SNR in
\eqref{Eq:SNR:DLIT}, the sum throughput is written as
\begin{equation}
\tilde{C}_f \!=\! \log_2(1+\theta)\! \sum_{n=1}^K\!  I\!\l( \frac{\beta_n P_nh_n }{\beta_n\sigma_a^2 + \sigma_b^2} \geq \theta \r)\!\! I(\!(1-\beta_n) P_n h_n \geq p_c) \nn
\end{equation}
that differs from the single-user counterpart in
\eqref{Eq:Rate:DL:SU:Share} by having the multi-user circuit-power
constraints. The matching power-control problem can be formulated as
\begin{equation}
\text{({\bf P6})}  \begin{aligned}
\underset{\{\beta_n, P_n\}}{\text{max}}\ \ & \sum_{n=1}^KI\!\l(\frac{\beta_n P_nh_n }{\beta_n\sigma_a^2 + \sigma_b^2}\geq \theta \r)\! I\!\l((1-\beta_n) P_n h_n \geq p_c\r)\\
\text{s.t.} \ \
& \sum\nolimits_{n=1}^K P_n \leq p_t,\\
&P_n \in\{0\}\cup \l[p_c/h_n, \infty\r)\quad \forall \ n,  \\
& \beta_n \in [0, 1]\quad \forall \ n.
\end{aligned}\nn
\end{equation}

\subsubsection{Solution} Replacing the inequalities in  P6 with
equalities has no effect on the solution. Hence, P6 can be rewritten
as
\begin{equation}
\text{({\bf P6})}  \begin{aligned}
\underset{\{\beta_n, P_n\}}{\text{max}}\ \ & \sum_{n=1}^KI\!\l(\frac{\beta_n P_nh_n }{\beta_n\sigma_a^2 + \sigma_b^2}= \theta \r)\! I\!\l((1-\beta_n) P_n h_n = p_c\r)\\
\text{s.t.} \ \
    & \sum\nolimits_{n=1}^K P_n = p_t,\\
        &P_n \geq 0 \ \forall \ n.
\end{aligned}\nn
\end{equation}
The splitting ratio for the $n$-th mobile can be obtained by solving
the following two linear equations:
\begin{equation}
\frac{\beta_n P_nh_n }{\beta_n\sigma_a^2 + \sigma_b^2}= \theta, \qquad (1-\beta_n) P_n h_n = p_c. \nn
\end{equation}
The resulting optimal value of $\{\beta_n\}$, which is identical for
all mobiles and denoted as $\beta^*$, has a similar form as the
single-user counterpart in \eqref{Eq:Beta:Op}:
\begin{equation}\label{Eq:Beta:Mu}
\beta_n = \tilde{\beta}^* = \frac{c(1) + \sqrt{c^2(1) + 4d}}{2}, \qquad \forall \ n
\end{equation}
where the coefficients $c(1)$ and $d$ are as given in
\eqref{Eq:Beta:Eq:Coeff}. With $\{\beta_n\}$ fixed as given in
\eqref{Eq:Beta:Mu}, it follows from inspecting P6 that the optimal
power-control policy again performs greedy channel inversion, just
like its single-user counterpart. The result is summarized in the
following proposition.
\begin{proposition}\emph{For the multi-user SWIPT system with downlink
    IT, the optimal power-control policy, represented by $\{P^*_n\}$,
    is given as
\begin{equation}
\l[P^*_1, P^*_2, \cdots, P^*_K\r]^T = \boldsymbol{\Pi}_h^{-1}\l[\bar{P}^*_1, \bar{P}^*_2, \cdots, P^*_{K}\r]^T
\end{equation}
where 
\begin{equation}
\bar{P}^*_n = \l\{
\begin{aligned}
&\frac{p_c}{(1-\tilde{\beta}^*) \bar{h}_n}, && 1\leq n \leq m_{\max}\\
&0, && \text{otherwise}.  
\end{aligned}
\r.
\end{equation}
The optimal splitting ratio $\tilde{\beta}^*$ is given by \eqref{Eq:Beta:Mu}
and $m_{\max}$, $1\leq m_{\max}\leq K$, is the largest integer
such that the power constraint
\begin{equation}
\frac{p_c}{1-\tilde{\beta}^*}\sum\nolimits_{n=1}^{m_{\max}}\frac{1}{\bar{h}_n} \leq p_t
\end{equation}
is satisfied.
}
\end{proposition}

\section{Power Control for Multi-User SWIPT Systems with Uplink IT}\label{Section:MU:ULIT}

\subsection{Multi-User Uplink IT with Variable Coding Rates}

\subsubsection{Problem formulation} The sum throughput for the current
case is given as
\begin{equation}
\tilde{R}_v = \sum\nolimits_{n=1}^K \log_2\l(1 + (P_ng'_n - p_c) g_n\r)I(P_n g'_n  \geq p_c). \label{Eq:Rate:UL}
\end{equation}
Note that the product $g_ng'_n$ in \eqref{Eq:Rate:UL} represents the
combined loss due to propagation both in the downlink and in the uplink.
This must be contrasted with the loss of only $h_n$ in the case of
downlink IT [see \eqref{Eq:Rate:DL:MU}].  The power-control problem is
formulated using \eqref{Eq:Rate:UL} as
\begin{equation}
\text{({\bf P7})} \quad \begin{aligned}
    \underset{\{P_n\}}{\text{max}}\quad &  \sum\nolimits_{n=1}^K \log\l(1 + (P_ng'_n - p_c) g_n\r) I(P_n g'_n \geq p_c)\\
\text{s.t.} \quad 
& \sum\nolimits_{n=1}^K P_n \leq p_t, \\
&P_n \in  \{0\}\cup\l(\frac{p_c}{g_n'}, \infty\r) \ \forall \ n. 
\end{aligned}\nn
\end{equation}

\subsubsection{Solution} To facilitate a compact exposition,  we use the following definitions. 
Let $\bar{g}'_1, \bar{g}'_2, \cdots, \bar{g}'_K$ denote
the downlink sub-channel gains $\{g'_n\}$ sorted in
 descending order and let  $\boldsymbol{\Pi}'_g$ be the corresponding permutation matrix; that is, we have:
\begin{equation}
[\bar{g}'_1, \bar{g}'_2, \cdots, \bar{g}'_K]^T = \boldsymbol{\Pi}'_g \times [g'_1, g'_2, \cdots, g'_K]^T. 
\end{equation}
Arranging the uplink sub-channel gains
 $\{g_n\}$ in the same way, i.e., $\bar{g}_1, \bar{g}_2, \cdots, \bar{g}_K$, gives
\begin{equation}
[\hat{g}_1, \hat{g}_2, \cdots, \hat{g}_K]^T = \boldsymbol{\Pi}'_g \times [g_1, g_2, \cdots, g_K]^T. 
\end{equation}
The powers $\{\bar{P}_n\}$ are defined based on $\{P_n\}$ in a similar way. 
Using these definitions, P7 can be rewritten as 
\begin{equation}
\text{({\bf P7.1})} \quad \begin{aligned}
    \underset{\{\bar{P}_n\}}{\text{max}}\quad &  \sum_{n=1}^K \log\l(1 + (\bar{P}_n\bar{g}'_n - p_c) \hat{g}_n\r) I(\bar{P}_n \bar{g}'_n \geq p_c)\\
\text{s.t.} \quad 
& \sum\nolimits_{n=1}^K \bar{P}_n \leq p_t, \\
&\bar{P}_n \in  \{0\}\cup\l(\frac{p_c}{\bar{g}_n'}, \infty\r) \ \forall \ n. 
\end{aligned}\nn
\end{equation}

Given that P7.1 is non-convex, a sub-optimal algorithm is proposed as follows. Assume that $k$
mobiles are active, that is, they harvest sufficient energy for
meeting their circuit-power constraints; all others are allocated zero
power. To maximize the MPT efficiency, the $k$ active mobiles are
chosen to be those corresponding to the largest downlink sub-channel
gains $\bar{g}'_1, \bar{g}'_2, \cdots, \bar{g}'_k$. This choice may not
be overall optimal, however, since selecting a mobile with relative
small downlink but sufficiently large uplink sub-channel gains can increase the
throughput.  Define $\bar{U}_n = \bar{P}_n - p_c/\bar{g}'_n$. Given
the assumptions and choices made, the problem of maximizing the uplink
sum throughput reduces to the standard multi-channel power
control problem:
\begin{equation}
\text{({\bf P7.1})} \quad \begin{aligned}
    \underset{\{\bar{U}_n\}}{\text{max}}\ \ &  \sum\nolimits_{n=1}^k \log\l(1 + \bar{U}_n \hat{g}_n\bar{g}'_n\r) \\
\text{s.t.} \quad 
& \sum\nolimits_{n=1}^k \bar{U}_n \leq p_t - p_c \sum\nolimits_{n=1}^k \frac{1}{\bar{g}'_n}, \\
&\bar{U}_n \geq 0  \ \forall \ 1\leq n\leq k. 
\end{aligned}\nn
\end{equation}
This problem is solved by water-filling:
\begin{equation}
\! \bar{U}^*_n(k) = \l\{
\begin{aligned}
&\frac{1}{k}\l(p_t - p_c\sum\nolimits_{n=1}^k \frac{1}{\bar{g}'_n }+\right.\\
&\quad \left. \sum\nolimits_{n=1}^k \frac{1}{\hat{g}_n\bar{g}'_n}\r) - \frac{1}{\hat{g}_n\bar{g}'_n}, &&1\leq n\leq k\\
&0, &&\text{otherwise}.  
\end{aligned}
\r.\label{Eq:U}
\end{equation}
The number of active mobile $k$ is optimized.  Let $z_{\max}$,
$1\leq z_{\max}\leq K$, be the maximum number of active mobiles such
that the corresponding multi-user circuit-power constraints and the
power constraint are satisfied:
\begin{equation}
p_c\sum\nolimits_{n=1}^{z_{\max}} \frac{1}{\bar{g}'_n} \leq p_t. 
\end{equation} 
For the same reason as discussed when solving P5.4, it may not be optimal to set
the optimal value of $k$, denoted as $k^*$, as $k^* = z_{\max}$.  Instead,
$k^*$ can be found by testing the values $1, 2, \cdots, z_{\max}$. The
above results are summarized in the following algorithm for computing
a sub-optimal solution for P7.
\begin{algorithm}\label{Algo:MU:ULIT:Var}\emph{
\begin{enumerate}
\item Compute the maximum number of active mobiles $z_{\max}$.
\item Determine the optimal number of streams $k^*$ as
\begin{equation}
k^* = \arg\max_{1\leq k \leq z_{\max}} \sum\nolimits_{n=1}^ k \log_2(1 + \bar{U}^*_n(k)) 
\end{equation}
with $\bar{U}^*_n(k)$ given in \eqref{Eq:U}. 
\item Given $k^*$, the allocated powers are computed as 
\begin{equation}
\bar{P}^*_n = \l\{
\begin{aligned}
& \bar{U}^*_n + \frac{p_c}{\bar{g}'_n}, && 1\leq n \leq k^*\\
& 0, && \text{otherwise}. 
\end{aligned}
\r.
\end{equation}
Then rearrange $\{\bar{P}^*_n\}$ to give the power allocation
$\{P^*_n\}$:
\begin{equation}
[P^*_1, P^*_2, \cdots, P^*_K]^T = (\boldsymbol{\Pi}'_g)^{-1} \times [\bar{P}^*_1, \bar{P}^*_2, \cdots, \bar{P}^*_K]^T. \nn
\end{equation}
\end{enumerate}}
\end{algorithm}

Algorithm~$1$ sequentially performs  the tasks of scheduling mobiles with high MPT efficiencies and maximizing the uplink sum rate of the scheduled mobiles by power control. The design exploits the fact that meeting the circuit power constraints   is a prerequisite for IT and hence has high priority. Such a sequential algorithm provides  a close-to-optimal solution as shown by simulation results in the sequel.

\subsection{Multi-User Uplink IT with Fixed Coding Rates}

\subsubsection{Problem formulation} The sum throughput for the current
scenario can be written as
\begin{equation}
\tilde{R}_f = \log_2(1+\theta) \sum\nolimits_{n=1}^K  I\l( Q_ng_n \geq \theta \r)I(P_n g_n \geq p_c).  \label{Eq:Rate:UL:MU:Share}
\end{equation}
The corresponding formulation of the optimal power-control problem follows as
\begin{equation}
\text{({\bf P8})} \quad \begin{aligned}
    \underset{\{P_n\}}{\text{max}}\quad & \sum\nolimits_{n=1}^KI\l((P_ng'_n - p_c)g_n \geq \theta \r)I\l(P_n g'_n \geq p_c\r)\\
\text{s.t.} \quad 
& \sum\nolimits_{n=1}^K P_n \leq p_t,\\
&P_n \geq 0 \ \forall \ n. 
\end{aligned}\nn
\end{equation}

\subsubsection{Solution}
Since the the first indicator function in the objective function of P8
yields $1$ if and only if the second does so, P8 reduces to
\begin{equation}
\text{({\bf P8})} \quad \begin{aligned}
\text{max}\quad & \sum\nolimits_{n=1}^KI\l( P_n \geq \frac{1}{g'_n}\l(\frac{\theta}{g_n} + p_c\r) \r)\\
\text{s.t.} \quad 
&P_n \geq 0 \ \forall \ n, \\
& \sum\nolimits_{n=1}^K P_n \leq p_t. 
\end{aligned}\nn
\end{equation}
For ease of notation, define the scalar sequence $v_1, v_2, \cdots, v_K$ according to
\begin{equation}\label{Eq:V:n}
v_n = \frac{1}{g'_n}\l(\frac{\theta}{g_n} + p_c\r) 
\end{equation}
and the vector $\bv = [v_1, v_2, \cdots, v_K]$. Let $\bar{v}_1,
\bar{v}_2, \cdots, \bar{v}_K$ represent the sequence $\{v_n\}$ sorted
in ascending  order, and define the vector $\bar{\bv} = [\bar{v}_1,
\bar{v}_2, \cdots, \bar{v}_K]$ and the permutation matrix
$\boldsymbol{\Pi}_v$ such that $\bar{\bv} = \boldsymbol{\Pi}_v\bv$. By
inspecting P8, the optimal power control policy at the base station is
found to be the one that attempts to meet the minimum-SNR requirements
of the uplink streams following the descending order of $\{v_n\}$. To
be specific, the optimal power allocated to the sub-channel
corresponding to $\bar{v}_n$, denoted as $\bar{P}_n$, is given as
\begin{equation}\label{Eq:Pwr:MU:UL:Fix}
\bar{P}_n^* = \left\{
\begin{aligned}
&\bar{v}_n, &&  1\leq n \leq q_{\max}\\
& 0, && \text{otherwise}
\end{aligned}\right.
\end{equation}
where $q_{\max}$ is the maximum number of uplink streams or
equivalently the largest integer for which  the power constraint
obtained from \eqref{Eq:PeakPwr:Const}, 
\begin{equation}
\sum\nolimits_{n=1}^{q_{\max}}\bar{P}_n^*\leq p_t,
\end{equation}
is satisfied.  Note that the policy $\{\bar{P}_n^*\}$ as specified by
\eqref{Eq:Pwr:MU:UL:Fix} is a variant of greedy channel inversion
where $\{\bar{v}_n\}$ combines the inversion of closed-loop channels
and circuit-power consumption.  Then the solution $\{P^*_n\}$ to P8
follows from rearranging $\{\bar{P}_n^*\}$ according to the original
order of the sub-channels. In other words,
\begin{equation}\label{Eq:Pwr:MU:UL:Fix:a}
[P_1^*, P_2^*, \cdots, P^*_K]^T = \boldsymbol{\Pi}_v^{-1}[\bar{P}_1^*, \bar{P}_2^*, \cdots, \bar{P}^*_K]^T
\end{equation}
with $\{\bar{P}^*_n\}$ in \eqref{Eq:Pwr:MU:UL:Fix}.  The key
results of this section are summarized in the following
proposition.
\begin{proposition}\emph{Consider the multi-user SWIPT system with
    uplink IT and fixed coding rates.
\begin{enumerate}
\item The optimal power-control policy at the base station is given by \eqref{Eq:Pwr:MU:UL:Fix:a}. 
\item It is optimal for each active mobile harvesting nonzero power to
  apply all available power for uplink transmission after deducting
the power needed to operate its circuitry.
\end{enumerate}
}
\end{proposition}

\section{Simulation Results}\label{Section:Simulation}
In this section, the performance of SWIPT using the power-control algorithms
proposed in the preceding sections is  evaluated by simulation in
terms of spectral efficiency  versus circuit power.  The channel model is described as follows. Propagation is assumed to have line-of-sight and be close to that in free space, which is necessary for making MPT feasible. The propagation model for beamed  transmission is modified from that in  \cite{Brown:BeamedMicrowavTarnsmsision} and specified by the following relation between the transmission power $P_t$ and received power $P_r$ for an arbitrary link:
\begin{equation}\label{Eq:Propagation}
\frac{P_r}{P_t} = \frac{A_t A_r }{\lambda^2 r^2}|Z|^2
\end{equation}
where $\lambda$ is the wavelength, $A_t$ and $A_r$ the total apertures of the transmit and receive antenna arrays, respectively, $r$ the transmission distance and $Z$ a complex Gaussian random variable with nonzero mean that models small-scale fading. For simulation, it is assumed that the wavelength corresponds to  a carrier frequency of $5.8$ GHz, the total aperture of the base-station antenna array is $1$ sq. m, the aperture of each of the two antennas at a mobile is $0.05$ sq. m, the base-station transmission power is $10$ W in the single-user system and $20$ W in the multi-user system, and $Z$ follows the $\mathcal{CN}(1, 0.2)$ distribution. For the scenario of uplink IT, the two sub-arrays at the base station that support full-duplex MPT/IT are assumed to have equal apertures of $0.5$ sq. m. For efficient MPT, transmission distances are assumed to be short as enabled by dense base-station deployment. To be specific, the distances are $100$ m for the single-user system and $\{50, 80, 100, 150, 200\}$ m for the multi-user system with five  mobiles. Correspondingly, there are five frequency  sub-channels which are assumed to be frequency non-selective. Their bandwidth has no effect on the simulation results since the performance metric is spectral efficiency. The distributions of the channel coefficients $\{\dot{h}_n, \ddot{h}_n, g_n, g_n'\}$ follow from the propagation model in \eqref{Eq:Propagation}. To be specific, each coefficient is given by the expression of $P_r/P_t$ in \eqref{Eq:Propagation} substituted with the corresponding transmission distance,  and all coefficients are assumed to be independent. 
Note that the beamforming gains are accounted for in the model of the channel coefficients via the antenna apertures \cite{Brown:BeamedMicrowavTarnsmsision}. Given short propagation distances  and line-of-sight channels, a mobile in can be exposed to extremely strong interference and hence the  interference-plus-noise variance    from each sub-channel is chosen to have a large value, namely $-30$ dBm, where $90\%$ and $10\%$ of the noise power  are  introduced before  and after a power splitter (see Fig.~\ref{Fig:Transceiver}), respectively. Note that in an interference dominant environment, the interference-plus-noise variance is largely determined by the ratio between the main-lobe and side-lobe responses rather than the channel bandwidth that affects the thermal noise variance. The  SNR threshold for the case of a fixed coding rate is set as $30$ dB for the scenario of downlink IT and $7$ dB for the scenario of uplink IT, which are optimized numerically to enhance the spectral efficiency. Last, the battery capacity at all mobiles is assumed to be sufficiently large such that  there is no energy loss due to battery overflow. 

\begin{figure*}
\begin{center}
\subfigure{\includegraphics[width=8cm]{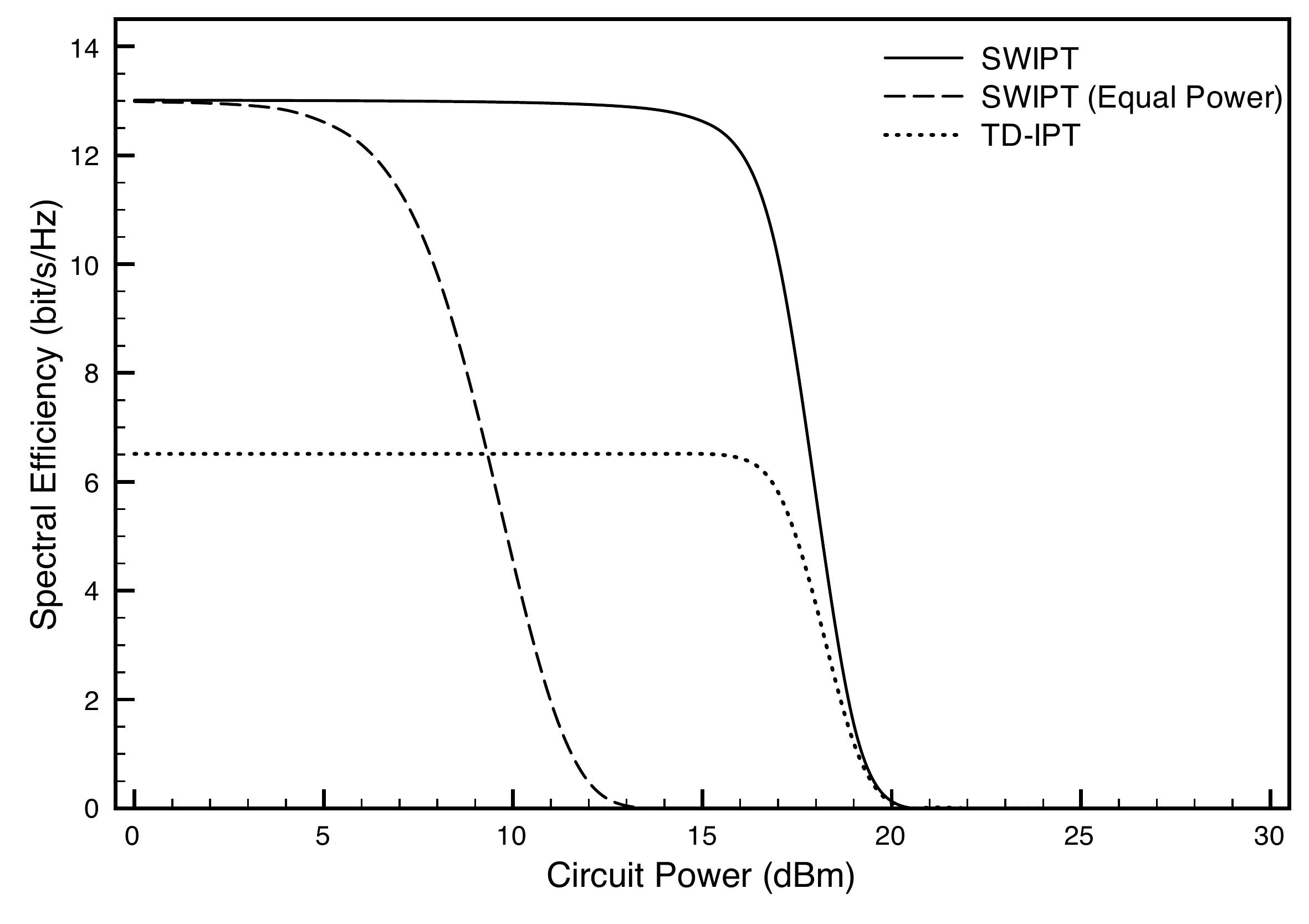}}
\subfigure{\includegraphics[width=8cm]{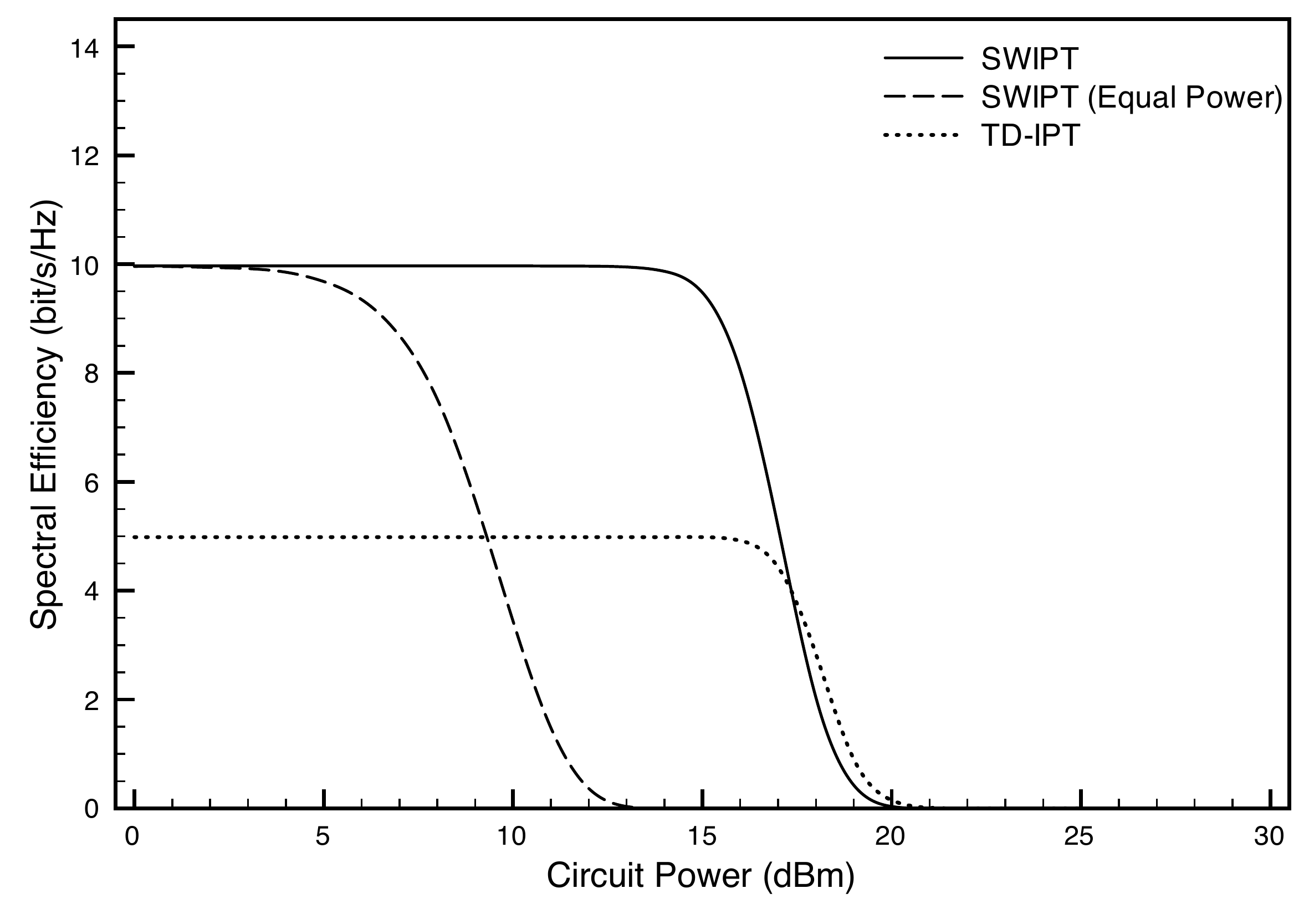}}\\
\subfigure{\includegraphics[width=8cm]{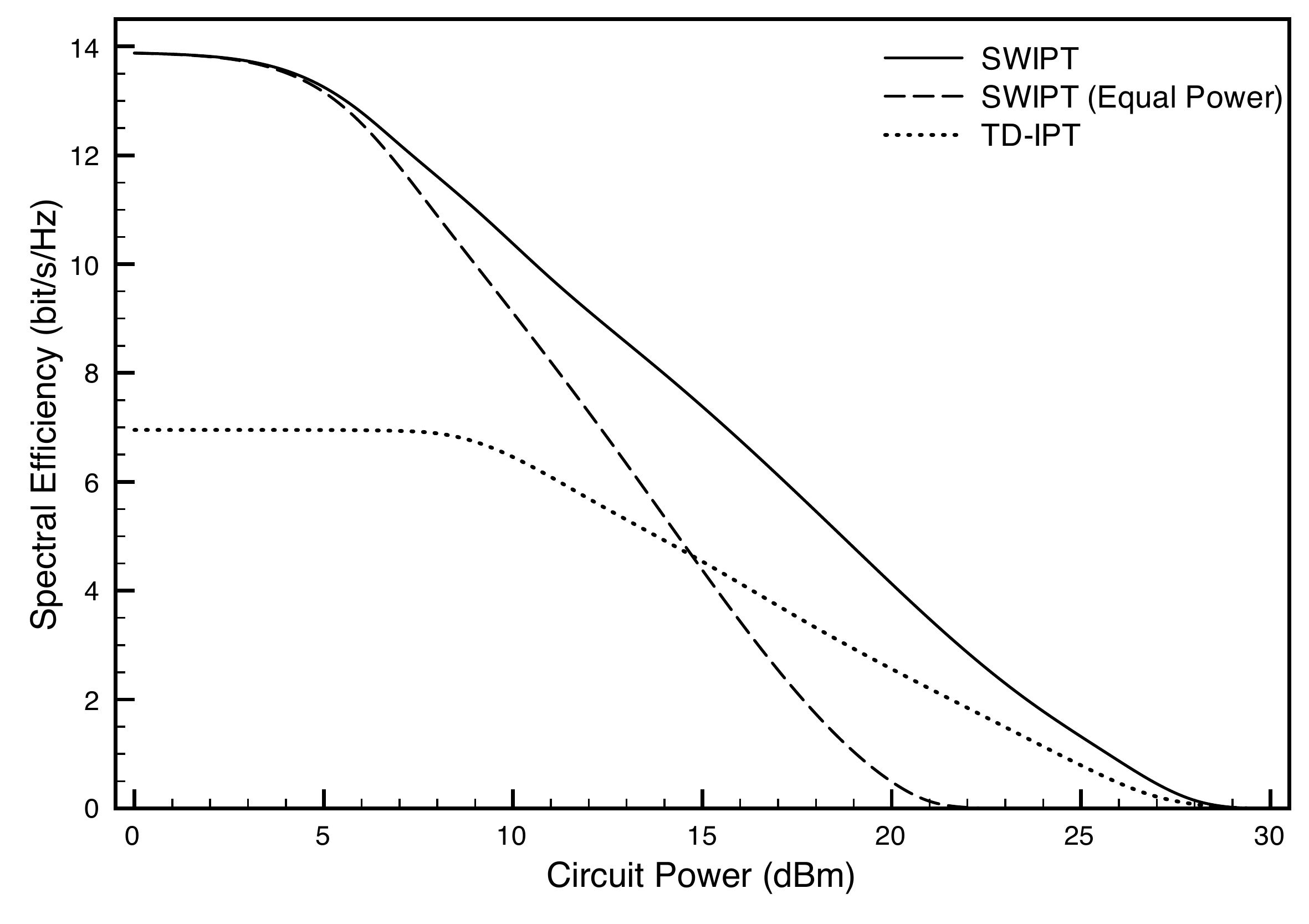}}
\subfigure{\includegraphics[width=8cm]{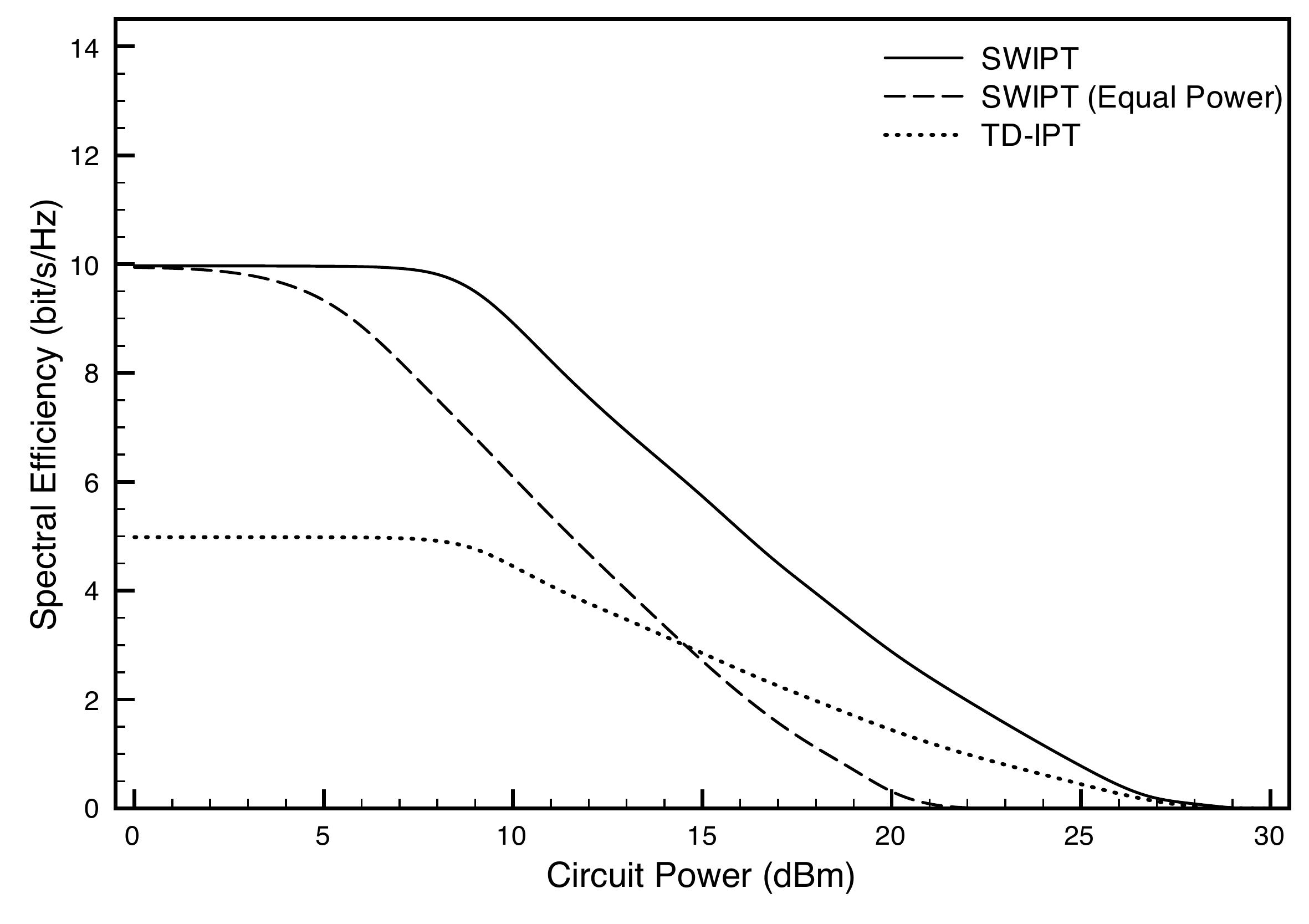}}
\caption{Spectral efficiency  versus circuit power for the scenario  of downlink IT.  (top left) Single user and variable coding rates. (top right) Single user and fixed coding rates. (bottom left) Multi-user  and variable coding rates. (bottom right) Multi-user and fixed coding rates.}
\label{Fig:SU:DLIT}
\end{center}
\end{figure*}

The proposed SWIPT with power control is compared in the sequel with SWIPT without such control (equal power allocation) as well as the  TD-IPT method  \cite{Zhang:MIMOBCWirelessInfoPowerTransfer, NgLo:MultiuserOFDMSInfoPowerTransfer}. It is assumed for TD-IPT that each time slot is divided into two halves for alternating  MPT and IT. The time sharing reduces the duration for IT by half but enhances the  received signal power by dedicating all antennas to  either MPT or IT at each  time instant. The power control algorithms for TD-IPT follow straightforwardly from those designed for SWIPT and thus the details are omitted for brevity.

\begin{figure*}
\begin{center}
\subfigure{\includegraphics[width=8cm]{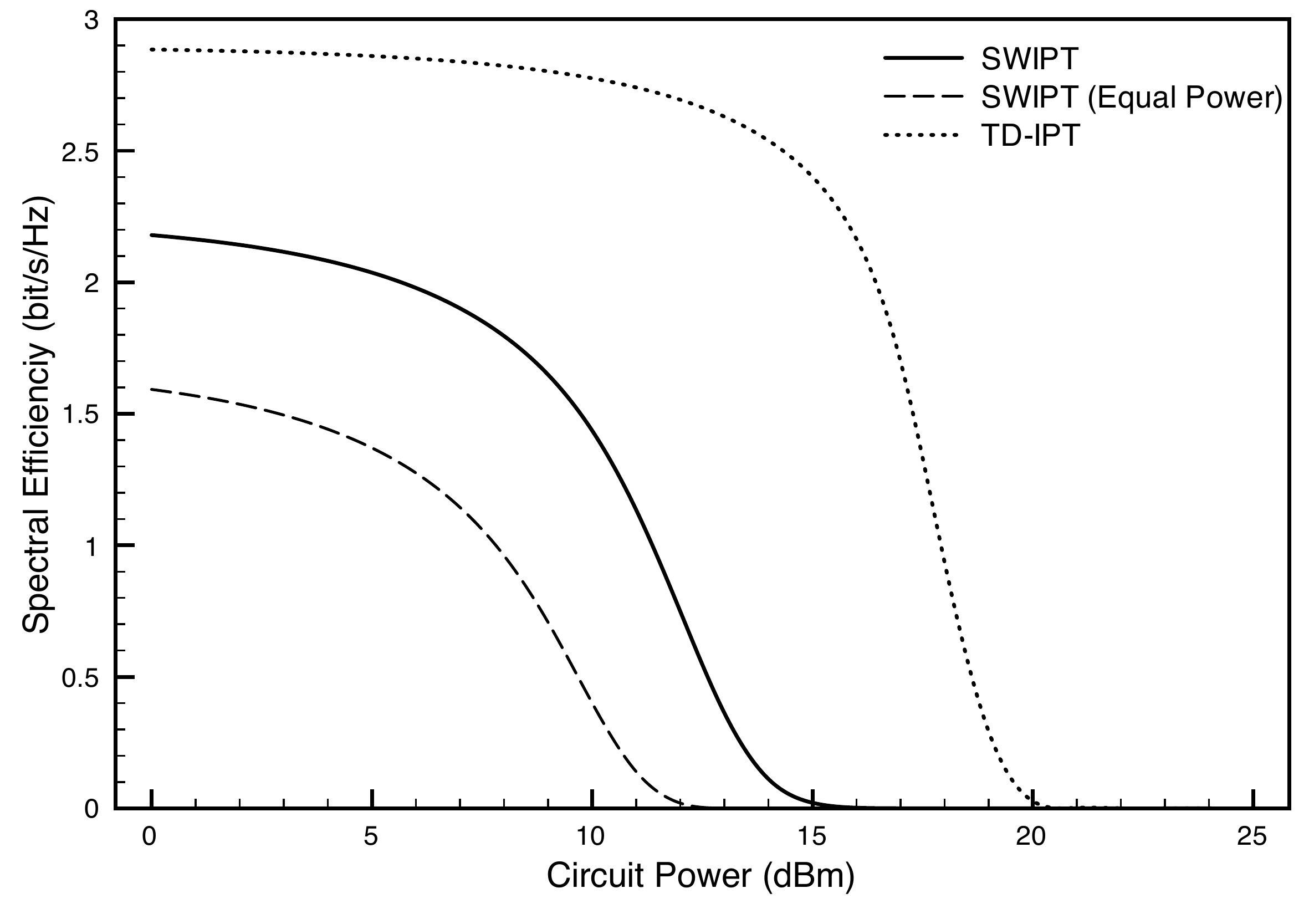}}
\subfigure{\includegraphics[width=8cm]{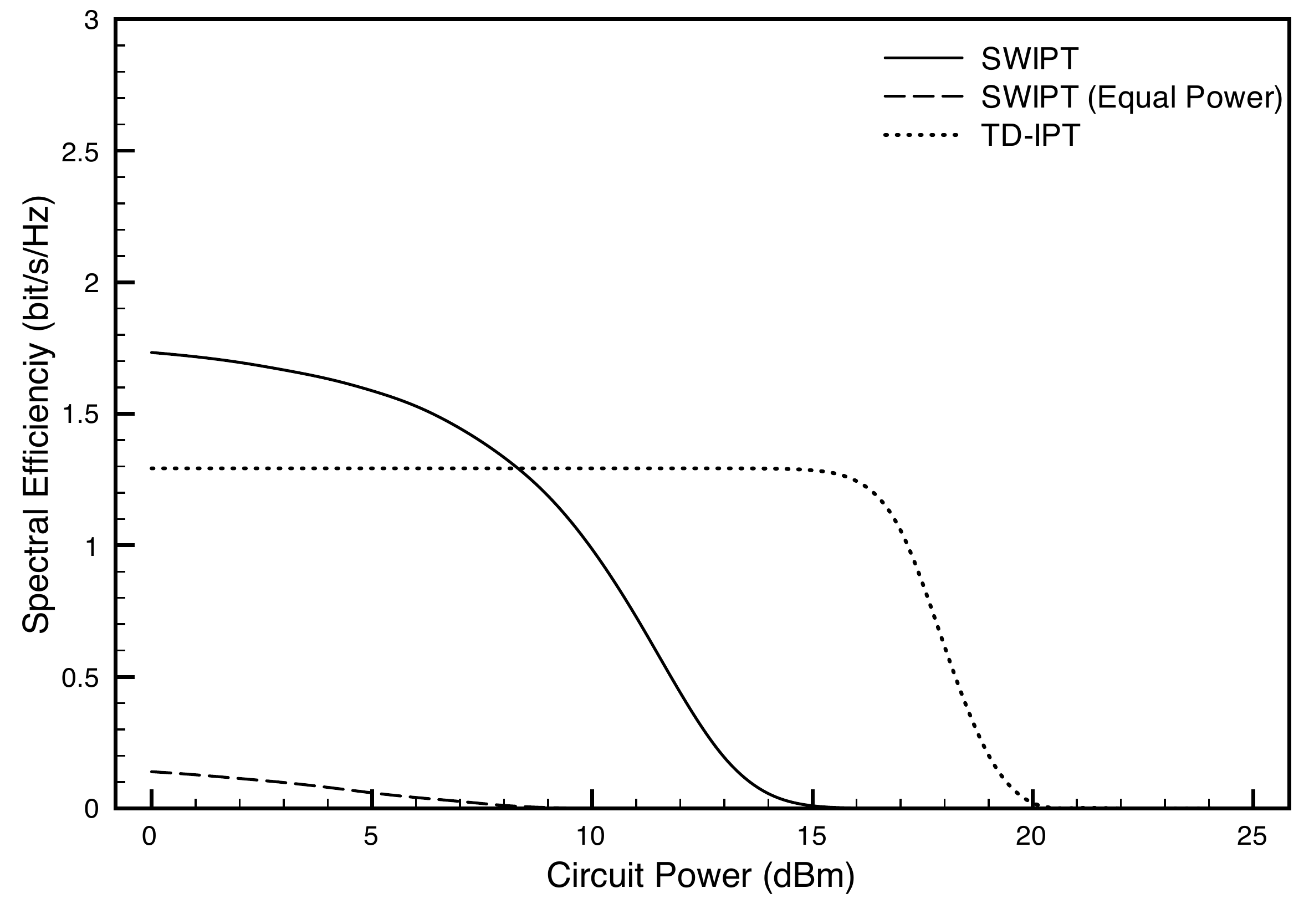}}\\
\subfigure{\includegraphics[width=8cm]{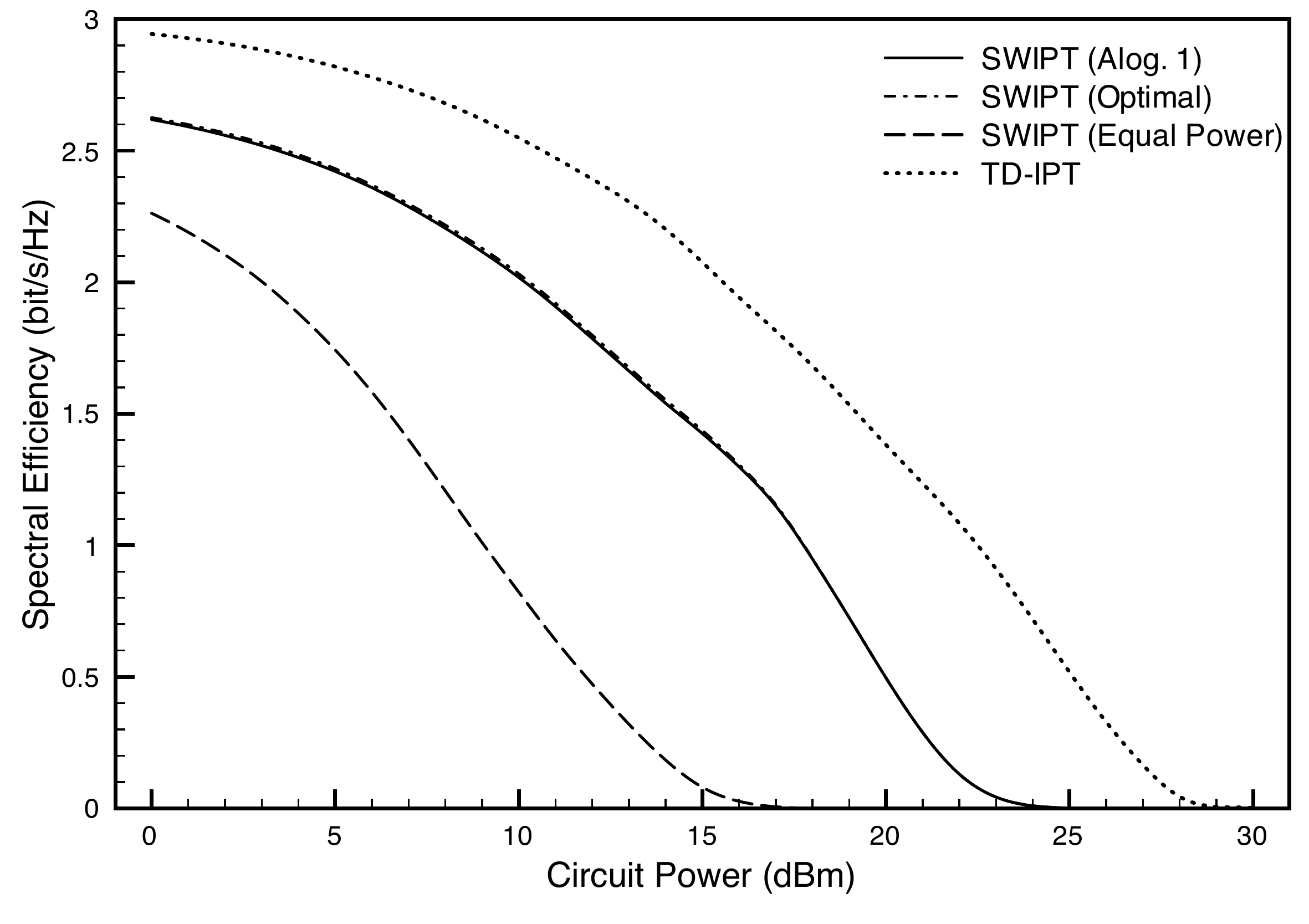}}
\subfigure{\includegraphics[width=8cm]{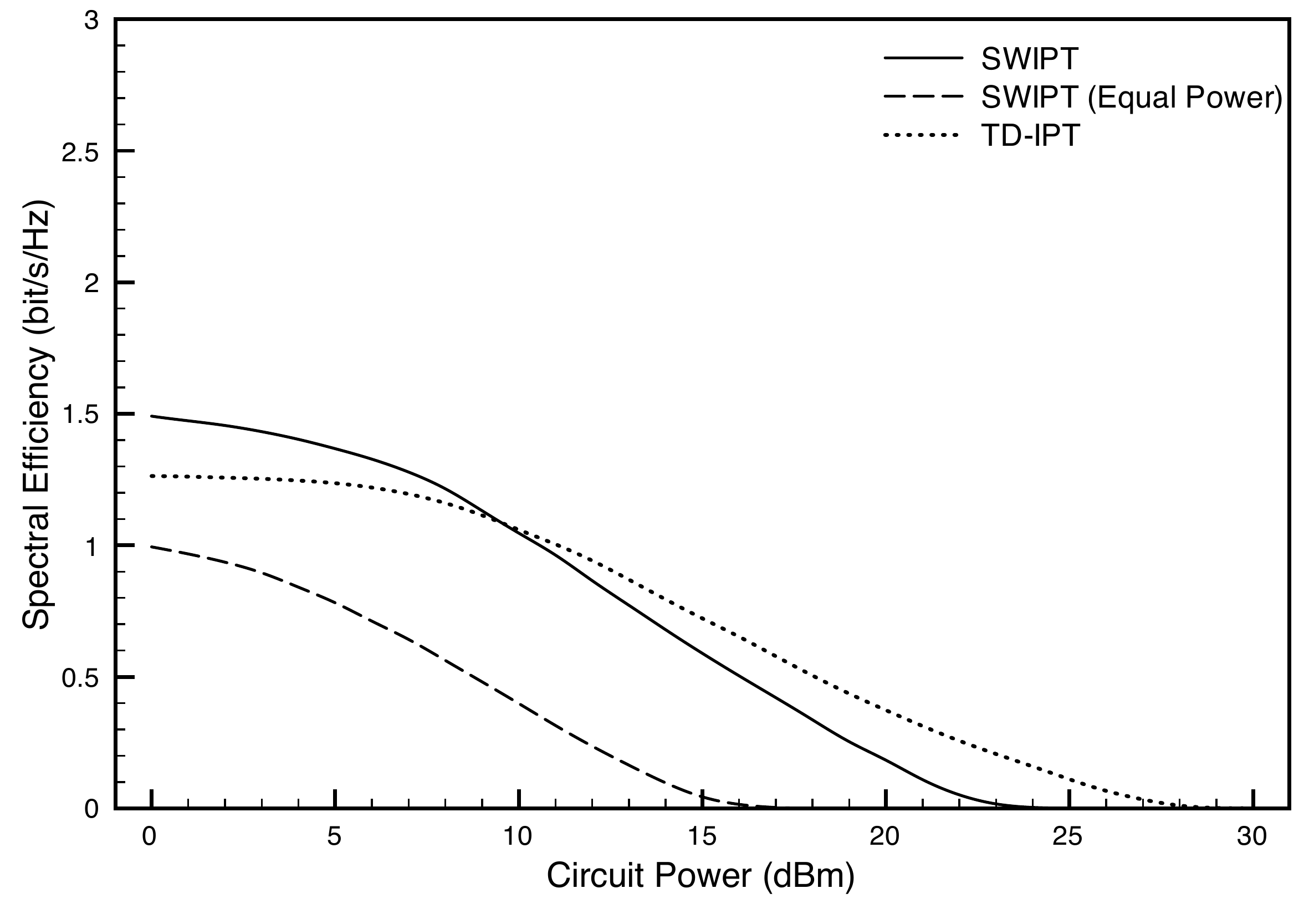}}
\caption{Spectral efficiency  versus circuit power for the scenario  of uplink  IT.  (top left) Single user and variable coding rates. (top right) Single user and  fixed coding rates. (bottom left) Multi-user  and variable coding rates. (bottom right) Multi-user and fixed coding rates.}
\label{Fig:SU:ULIT}
\end{center}
\end{figure*}

First, consider the scenario of SWIPT with downlink IT. The curves of spectral efficiency versus circuit power are  plotted in the sub-figures in Fig.~\ref{Fig:SU:DLIT}, corresponding to different cases combining single-user/multi-user systems and variable/fixed coding rates.  For all the curves in
the  figure, as the circuit power decreases, the spectral efficiencies  converge 
to their counterparts for the case with reliable
power supplies at the mobiles, which are extremely high ($10-13$ bit/s/Hz)  due to the low propagation loss.  The spectral efficiencies   reduce with increasing circuit power. In particular, the changes exhibit a threshold effect for the single-user system (see top sub-figures in Fig.~\ref{Fig:SU:DLIT}). This suggests that  powering one  passive mobile by MPT has little effect on the spectral efficiency  if the circuit power is below the threshold, but otherwise it degrades the efficiency severely. However, for the multi-user system, since the base-station needs to power multiple mobiles, the spectral efficiency is sensitive to the changes in the circuit power (see bottom sub-figures in Fig.~\ref{Fig:SU:DLIT}). Next, comparing SWIPT with and without power control, it is observed that with the spectral efficiency fixed  such control can increase circuit power substantially e.g., by up to about $8$ dB for the single-user system. Last, though TD-IPT yields spectral efficiencies about half of those by SWIPT for low to moderate circuit power, the gap narrows as the power increases and TD-IPT can outperform SWIPT for high circuit power as shown in the case of the single-user system with fixed coding rates.

Next, consider the scenario of SWIPT with uplink IT.  A similar set of curves as those in Fig.~\ref{Fig:SU:DLIT} are plotted in Fig.~\ref{Fig:SU:ULIT}. Compared with the previous scenario of SWIPT with downlink IT, the power supplied by the base station must overcome a \emph{roundtrip} propagation loss, first for the MPT in the downlink and then for the IT in the uplink, which decreases the spectral efficiencies by more than $10$ bit/s/Hz. For the current scenario, TD-IPT is found to outperform SWIPT. This suggests that given severe propagation loss  it should be preferable to use all transmit/receive antennas for either MPT or IT which more than compensates the  time-sharing loss. Last, the performance of the sub-optimal Algorithm~$1$ designed for the case of multi-user SWIPT with uplink IT is observed to be close-to-optimal, where the curve for the optimal algorithm is obtained by scheduling based on  an exhaustive search for maximizing the spectral efficiency.

\begin{figure*}
\begin{center}
\subfigure{\includegraphics[width=8cm]{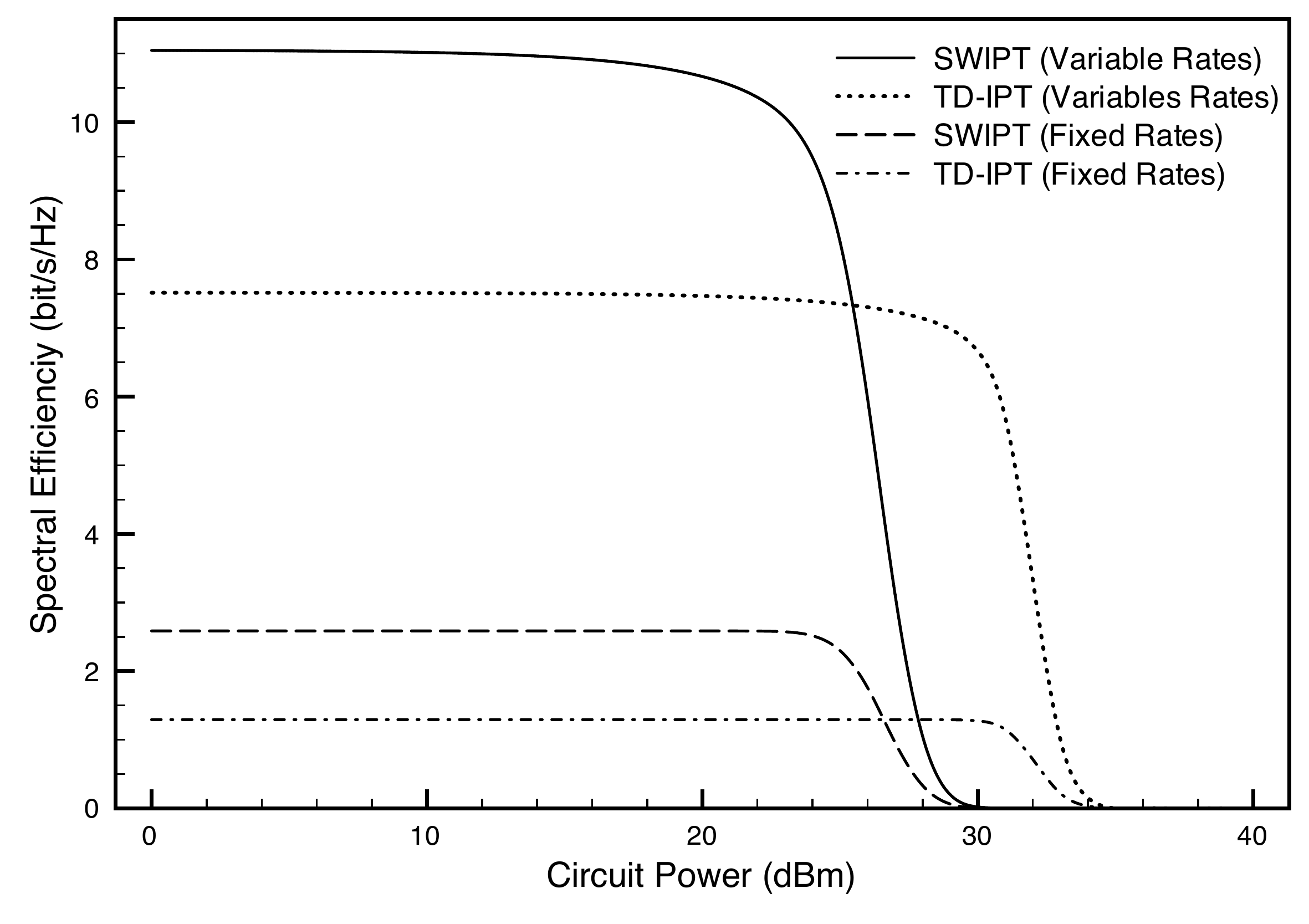}}
\subfigure{\includegraphics[width=8cm]{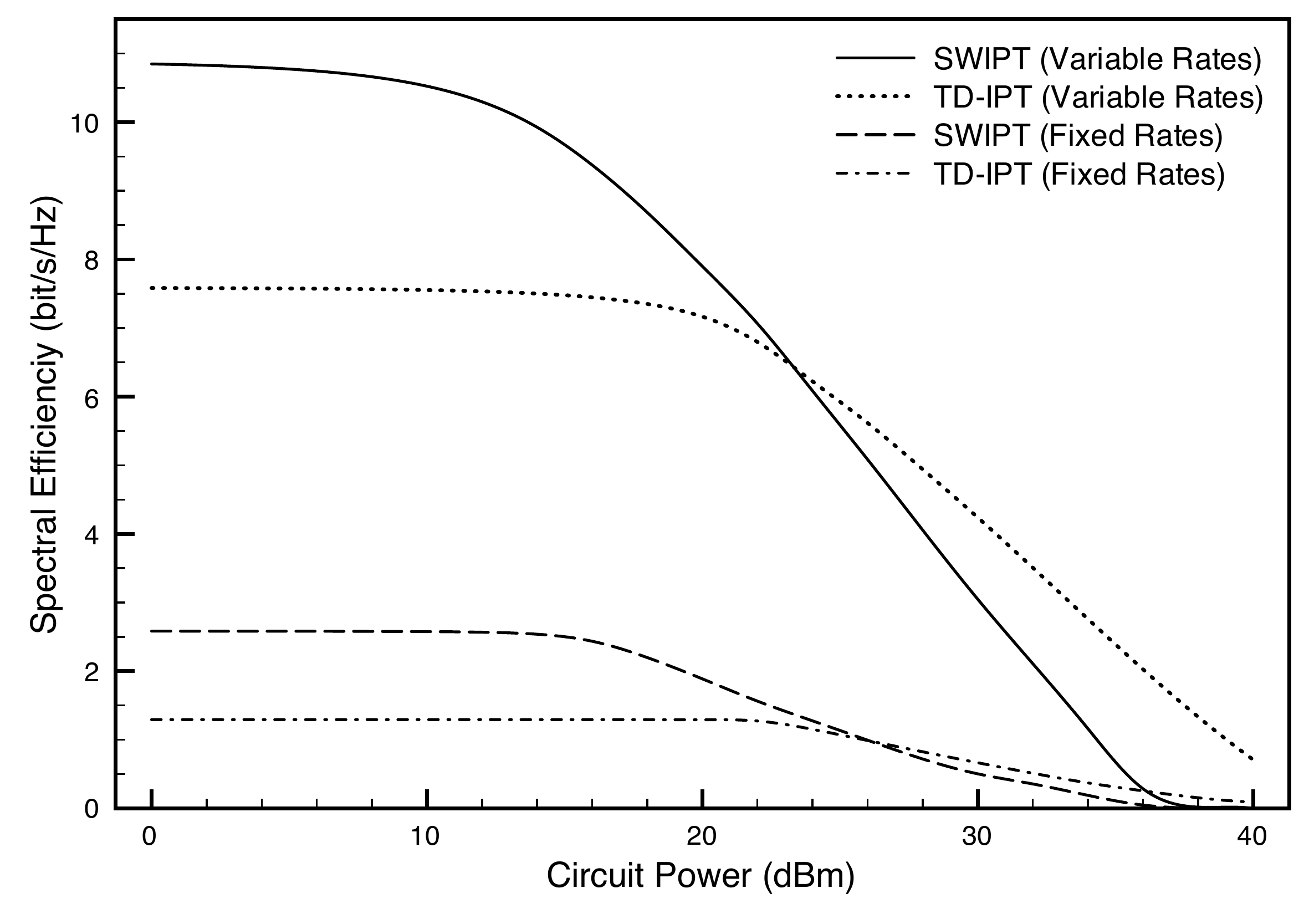}}
\caption{Spectral efficiency  versus circuit power for the scenario  of uplink IT with reduced transmission distances, namely $20$ m for the single-user system and $\{10, 16, 20, 30, 40\}$ m for the multi-user system. (left) Single user. (right) Multi-user. }
\label{Fig:TD-IPT:Compare}
\end{center}
\end{figure*}

For the same scenario of uplink IT, a further comparison between TD-IPT and SWIPT is provided in Fig.~\ref{Fig:TD-IPT:Compare} for which the round-trip propagation loss is alleviated by reducing all transmission distances by five times. It is observed that there are intersections between the curves for SWIPT and their TD-IPT counterparts. This leads to the conclusion that SWIPT is preferred when the propagation loss is not extremely severe (e.g., for the case of downlink IT) or the circuit power is low; otherwise, TD-IPT should be used for a higher spectral efficiency.

\section{Conclusions}\label{Section:Conclusion}

A framework has been proposed for realizing SWIPT in a broadband
wireless system that comprises a passive SWIPT-enabled mobile
architecture and a matching set of power-control algorithms designed
for different system configurations accounting for
single-user/multi-user systems, variable/fixed coding rates, and
uplink/downlink information transfer. These algorithms have been
optimized for maximizing the system throughput under circuit-power
constraints at mobiles, in addition to a power constraint at the
base station.  It is shown by simulation that power control plays an
important role in enhancing the efficiency of SWIPT.

This work can be extended in several interesting directions. First,
the channel assignment was assumed to be fixed here. Jointly assigning
channels and performing optimal power control may further increase the
SWIPT efficiency. Second, the current framework can be modified to
support multimode operations including MPT or SWIPT to nearby mobiles
but only information transfer to mobiles far away. Third, the power
control can be integrated with intelligent energy management policies
at the mobiles, in order to exploit the diversity that originates from
time-variations of the channels. Finally, it would be interesting to
design a framework for cooperative SWIPT in a multi-cell system.

\bibliographystyle{ieeetr}

\begin{IEEEbiography}[{\includegraphics[width=1in, clip, keepaspectratio]{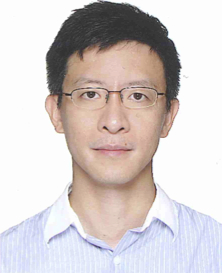}}]{Kaibin Huang}  (S'05, M'08, SM'13) received the B.Eng. (first-class hons.) and the M.Eng. from the National University of Singapore in 1998 and 2000, respectively, and the Ph.D. degree from The University of Texas at Austin (UT Austin) in 2008, all in electrical engineering.

Since Jul. 2012, he has been an assistant professor in the Dept. of Applied Mathematics (AMA) at The Hong Kong Polytechnic University (PolyU), Hong Kong. He had held the same position in the School of Electrical and Electronic Engineering at Yonsei University, S. Korea from Mar. 2009 to Jun. 2012 and presently is affiliated with the school as an adjunct professor. From Jun. 2008 to Feb. 2009, he was a Postdoctoral Research Fellow in the Department of Electrical and Computer Engineering at the Hong Kong University of Science and Technology. From Nov. 1999 to Jul. 2004, he was an Associate Scientist at the Institute for Infocomm Research in Singapore. He frequently serves on the technical program committees of major IEEE conferences in wireless communications. He will chair the Comm. Theory Symp. of IEEE GLOBECOM 2014 and the Adv. Topics in Wireless Comm. Symp. of IEEE/CIC ICCC 2014, and has been the technical co-chair for IEEE CTW 2013, the track chair for IEEE Asilomar 2011, and the track co-chair for IEE VTC Spring 2013 and IEEE WCNC 2011. He is a guest editor for the IEEE Journal on Selected Areas in Communications,  and an editor for the IEEE Wireless Communications Letters and also the Journal of Communication and Networks. He is an elected member of the SPCOM Technical Committee of the IEEE Signal Processing Society. Dr. Huang received the Outstanding Teaching Award from Yonsei, Motorola Partnerships in Research Grant, the University Continuing Fellowship at UT Austin, and Best Paper Awards from IEEE GLOBECOM 2006 and PolyU AMA. His research interests focus on the analysis and design of wireless networks using stochastic geometry and multi-antenna limited feedback techniques.
\end{IEEEbiography}

\begin{IEEEbiography}[{\includegraphics[width=1in, clip, keepaspectratio]{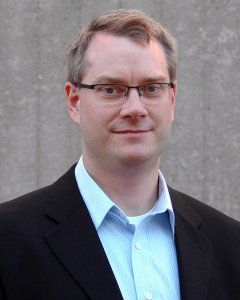}}]{Erik G. Larsson}   received his Ph.D. degree from Uppsala University,
Sweden, in 2002.  Since 2007, he is Professor and Head of the Division
for Communication Systems in the Department of Electrical Engineering
(ISY) at Link\"oping University (LiU) in Link\"oping, Sweden. He has
previously been Associate Professor (Docent) at the Royal Institute of
Technology (KTH) in Stockholm, Sweden, and Assistant Professor at the
University of Florida and the George Washington University, USA.

His main professional interests are within the areas of wireless
communications and signal processing. He has published some 100 journal papers
on these topics, he is co-author of the textbook \emph{Space-Time
Block Coding for Wireless Communications} (Cambridge Univ. Press,
2003) and he holds 10 patents on wireless technology.

He is Associate Editor for the \emph{IEEE Transactions on 
Communications} and he has previously been Associate Editor 
for several other IEEE journals. He is a member of
the IEEE Signal Processing Society  SPCOM technical
committee. He is active in conference organization, most recently as
the Technical Chair of the  Asilomar Conference on Signals, Systems 
and Computers 2012 and Technical Program co-chair of the
International Symposium on Turbo Codes and Iterative Information Processing 2012.
He received the \emph{IEEE Signal Processing Magazine} Best Column Award 2012.

\end{IEEEbiography}

\vfill

\end{document}